\documentclass[submission,copyright,creativecommons]{eptcs}

\providecommand{\event}{GandALF 2021}

\usepackage[noframe]{showframe}
\usepackage{wrapfig}
\usepackage{pgf,tikz}
\usepackage{breakurl}
\usepackage{underscore}
\usepackage{amssymb}
\usepackage{paralist}
\usepackage{refcount}
\usepackage{enumitem}
\usepackage{fdsymbol}
\usepackage{amsthm}

\usetikzlibrary{calc}
\usetikzlibrary{patterns}
\usetikzlibrary{backgrounds}
\usetikzlibrary{fit}
\usetikzlibrary{decorations.pathreplacing}
\pgfdeclarelayer{background}
\pgfdeclarelayer{foreground}
\pgfsetlayers{background,main,foreground}

\newcommand{\vpair}[1]{$\begin{array}{c}
   #1    
\end{array}$}
\newcommand{\hsER}{\langle R\rangle}
\newcommand{\hsEB}{\langle B\rangle}
\newcommand{\hsEA}{\langle A\rangle}

\newcommand{\hsED}{\langle D\rangle}
\newcommand{\hsEE}{\langle E\rangle}

\newcommand{\hsAR}{[R]}
\newcommand{\hsAA}{[A]}
\newcommand{\hsAB}{[B]}
\newcommand{\hsAD}{[D]}

\newcommand{\equivmodA}{\equiv_{\neg A}}
\newcommand{\nequivmodA}{\nequiv_{\neg A}}

\newcommand{\hsA}{[G]}

\renewcommand{\min}{min}

\newcommand{\LogicABDhom}{\ensuremath{\mathrm{ABD}}}

\newcommand{\Prop}{\ensuremath{\mathrm{Prop}}}

\newcommand{\TFA}{TF^\varphi_A}

\newcommand{\bfM}{\mathbf{M}}

\newcommand{\hx}{\hat{x}}
\newcommand{\hy}{\hat{y}}
\newcommand{\otile}{\overline{tile}}
\newcommand{\bbN}{\mathbb{N}}
\newcommand{\bbQ}{\mathbb{Q}}
\newcommand{\bbR}{\mathbb{R}}

\newcommand{\witnesses}{\mathrm{Wit}_{\cG}}

\newcommand{\bN}{\mathbb{N}}

\newcommand{\bS}{\mathbb{S}}

\newcommand{\cG}{\mathcal{G}}
\newcommand{\cL}{\mathcal{L}}

\newcommand{\cP}{\mathcal{P}}

\newcommand{\cT}{\mathcal{T}}
\newcommand{\cV}{\mathcal{V}}

\newcommand{\bI}{\mathbb{I}}
\newcommand{\bG}{\mathbb{G}}

\newcommand{\rowG}{\mathrm{Row}_{\cG}}

\renewcommand{\closure}{\mathrm{Cl}(\varphi)}
\newcommand{\atoms}{\mathrm{At}(\varphi)}

\newcommand{\ox}{\overline{x}}
\newcommand{\oy}{\overline{y}}

\newcommand{\fpG}{fp_{\cG}}

\newcommand{\thenB}{\rightarrow_B}
\newcommand{\thenD}{\rightarrow_D}

\newcommand{\shadingB}{\mathrm{Sh}_{B}}

\newcommand{\shadingG}{\mathrm{Sh}^{\cG}}
\newcommand{\shadingGB}{\shadingG_B}
\newcommand{\shadingGN}{\shadingG_{\bbN}}

\newcommand{\obsR}{\mathrm{Obs}_R}
\newcommand{\obsB}{\mathrm{Obs}_B}
\newcommand{\obsD}{\mathrm{Obs}_D}

\newcommand{\reqR}{\mathrm{Req}_R}
\newcommand{\reqB}{\mathrm{Req}_B}
\newcommand{\reqD}{\mathrm{Req}_D}
\newcommand{\reqA}{\mathrm{Req}_A}

\newcommand{\boxR}{\mathrm{Box}_R}
\newcommand{\boxB}{\mathrm{Box}_B}
\newcommand{\boxD}{\mathrm{Box}_D}
\newcommand{\boxA}{\mathrm{Box}_A}

\newcommand{\Deltareq}{\Delta_{\uparrow}}
\newcommand{\future}{\bS_{\rightarrow}}

\newcommand{\arraycellcomment}[3]{#1 \mbox{\begin{tabular}{p{#2}} #3 \end{tabular}}}

\newcommand{\tileH}{{\Rightarrow}}
\newcommand{\tileV}{{\Uparrow}}

\newcommand{\asat}{\medblacklozenge}
\newcommand{\areq}{\medlozenge}
\newcommand{\abox}{\medsquare}

\newtheorem{theorem}{Theorem}
\newtheorem{problem}{Problem}
\newtheorem{lemma}{Lemma}
\newtheorem{proposition}{Proposition}
\newtheorem{corollary}{Corollary}
\newtheorem{definition}{Definition}

\title{Adding the Relation \emph{Meets} to the Temporal Logic of \\ Prefixes and Infixes Makes It EXPSPACE-Complete}
\author{Laura Bozzelli
\institute{University Federico II,\\ Naples, Italy}
\institute{Department of Electric Engineering\\ and Information Technology}
\email{laura.bozzelli@unina.it}
\and
Angelo Montanari
\institute{University of Udine, Italy}
\institute{Department of Computer Science,\\ Mathematics, and Physics}
\email{angelo.montanari@uniud.it}
\and
Adriano Peron
\institute{University Federico II,\\ Naples, Italy}
\institute{Department of Electric Engineering\\ and Information Technology}
\email{adriano.peron2@unina.it}
\and
Pietro Sala
\institute{University of Verona, Italy}
\institute{Department of Computer Science}
\email{pietro.sala@univr.it}}
\def\titlerunning{Adding \emph{Meets} to the Temporal Logic 
	of Prefixes and Infixes
	Makes It EXPSPACE-Complete.}
\def\authorrunning{L. Bozzelli, A. Montanari, A. Peron \& P. Sala}
\begin{document}
\maketitle
\begin{abstract}
The choice of the right trade-off between expressiveness and complexity is the main issue in interval temporal logic. In their seminal paper~\cite{DBLP:journals/jacm/HalpernS91}, Halpern and Shoham showed that the satisfiability problem for $\mathsf{HS}$ (the temporal logic of Allen's relations) is highly undecidable over any reasonable class of linear orders. In order to recover decidability, one can restrict the set of temporal modalities and/or the class of models. In the following, we focus on the satisfiability problem for $\mathsf{HS}$ fragments under the homogeneity assumption, 
according to which any proposition letter holds over an interval if only if it holds at all its points. The problem for full $\mathsf{HS}_{hom}$ has been shown to be non-elementarily decidable~\cite{DBLP:journals/acta/MolinariMMPP16}, but its only known lower bound is EXPSPACE (in fact, EXPSPACE-hardness has been shown for the logic of prefixes and suffixes $\mathsf{BE}_{hom}$, which is a very small fragment of it~\cite{DBLP:journals/tcs/BozzelliMMPS19}). The logic of prefixes and infixes $\mathsf{BD}_{hom}$ has been recently shown to be PSPACE-complete \cite{BMPS21}. In this paper, we prove that the addition of the Allen relation \emph{Meets} to $\mathsf{BD}_{hom}$ makes it EXPSPACE-complete.

\end{abstract}

%!TEX root = ./main.tex 

\section{Introduction}\label{sec:intro}

Interval temporal logics (ITLs for short) are versatile and expressive formalisms for specifying properties of sequences of states and their durations. When it comes to fundamental problems like satisfiability, their high expressive power is often paid at the price of undecidability. For instance, the most widely known ITLs, that is, HS logic, proposed by Halpern and Shoham's \cite{DBLP:journals/jacm/HalpernS91}, and CDT logic, proposed by Venema \cite{10.1093/logcom/1.4.453}, turn out to be highly undecidable w.r.t.~the satisfiability problem. Despite these negative results, a number of decidable formalisms have been identified by weakening ITLs (see \cite{DBLP:journals/tcs/BresolinMMSS14} for a complete classification of HS fragments). 
Here the term ``weakening'' is intended as a set of syntactic and/or semantics restrictions imposed on the formulas of the logic
and/or the models on which such formulas are interpreted, respectively. 
Among the plethora of possible weakenings, in this paper 
we focus on (the combination of) the following two natural and well-studied restrictions:

\begin{itemize}

\item \textbf{Restrict the set of interval relations.} Many decidable fragments of ITLs are obtained by considering a restricted set of Allen's relations for comparing intervals. This approach naturally induces fragments of HS logic with modal operators corresponding to the selected subset of interval relations. As an example, the logic of temporal neighborhood, called PNL,  features only two interval relations among the possible 13 ones, namely, $A$ (\emph{adjacent to the right}) and its inverse $\bar A$. The corresponding interval modal logic has been shown to be decidable over basically every class of linear orders (e.g. see \cite{DBLP:conf/tableaux/BresolinMSS11,DBLP:conf/time/MontanariS12});

\item \textbf{Restrict the class of models.} Based on a  principle similar to the above one, some ITLs can be tamed by considering classes of models that satisfy certain specific assumptions. An example of this type of restriction can be found in a series of recent papers that studied model-checking problems for ITLs (e.g. see the seminal paper \cite{DBLP:journals/acta/MolinariMMPP16})
%,DBLP:journals/iandc/MolinariMP18
, as well as ITL expressiveness compared to classical point-based temporal logics, like LTL, CTL, and CTL$^*$ \cite{DBLP:journals/tocl/BozzelliMMPS19}. In this setting, models are represented as Kripke structures, and so inherently point-based rather than interval-based. The generated models can be equivalently obtained by making the so-called \emph{homogeneity assumption}, that is, by assuming that every proposition letter holds over an interval if and only if it holds at all its points. It is important noticing that, under the homogeneity assumption, the full HS logic has a decidable satisfiability problem (as a matter of fact, the model-checking procedures introduced in the aforementioned works can be easily turned to satisfiability procedures, while often retaining the same complexity). Because of this, the focus in studying HS logics under the homogeneity assumption is shifted from decidability to complexity. 
\end{itemize}

Let us focus now on the Chop logic 
$\mathsf{C}$ which is a proper 
fragment of $\mathsf{CDT}$
admitting as single  modality called \emph{chop} operator, that allows one to split the current interval in two parts and to require properties to hold separately on the two parts. In the general setting
even the satisfiability problem for the 
logic $\mathsf{C}$ is undecidable, however if we impose the homogeneity constraint for satisfiability, thus obtaining the logic
$\mathsf{C}_{hom}$, then it can be easily shown that there is a $LOGSPACE$ reduction of the emptiness problem for star-free generalized regular expressions to the satisfiability problem for $\mathsf{C}_{hom}$ and vice versa. However, a classic result in formal languages proved by Stockmeyer states that the emptiness problem for star-free generalized regular expressions is non-elementarily decidable 
(tower-complete) ~\cite{Schmitz:2016,stockmeyer1974complexity}. This means that the satisfiability 
problem for $\mathsf{C}_{hom}$ 
is also non-elementarily decidable. 

As we mentioned above, the 
satisfiability problem for the full logic $\mathsf{HS}$
when interpreted under the homogeneity assumption, called $\mathsf{HS}_{hom}$ from now on, is decidable \cite{DBLP:journals/acta/MolinariMMPP16},
and the only known decision 
algorithm for such problem features a non-elementary complexity. To this day, the exact complexity
of the satisfiability problem for $\mathsf{HS}_{hom}$ is still an open problem since a matching lower bound has not been fixed yet. 
From an expressivity standpoint, the aforementioned 
logic  $\mathsf{C}_{hom}$  
can capture only three  
of the $12$  
modal operators featured by $\mathsf{HS}$, namely, the operators $B$, for ``begins'', 
corresponding to the prefix relation on pairs of intervals, $D$, for
``during'', corresponding to the infix relation on pairs of intervals, and $E$, for ``ends'', 
corresponding to the suffix relation on pairs of intervals. It is easy to see that any fragment of $\mathsf{HS}_{hom}$ that contains both the 
operators $B$ and $E$
also contains $D$, since the infix
relation may be expressed, for istance, as some prefix of some suffix or vice versa. Let us notice that the opposite is not true, e.g., the 
$\mathsf{BD}_{hom}$ fragment of 
$\mathsf{HS}_{hom}$ cannot express 
the $E$ operator. Informally speaking, this means that  not all the properties about the prefixes of an interval cannot be expressed in terms prefixes and infixes. 
The satisfiability problem for 
the fragment $BE_{hom}$ has been proved to be EXPSPACE-hard 
in \cite{DBLP:journals/tcs/BozzelliMMPS19} while the upper bound ranges from $EXPSPACE$ to non-elementary. We conjecture that the complexity of the satisfiability problem of  
$BE_{hom}$ coincides with the complexity of the problem of the full logic $\mathsf{HS}_{hom}$.

A couple of arguments  
that make $\mathsf{BE}_{hom}$ such a 
peculiar beast are the following: (i) according to the results 
proved/summarized in \cite{DBLP:journals/tcs/BozzelliMMPS19} w.r.t. the satisfiability problem
the only known fragments of $HS_{hom}$ for which it has been possible to provide an EXPSPACE lower bound for the complexity must contain  both $B$ and
 $E$ operators;
 (ii) the satisfiability problem for 
 the logic $\mathsf{DE}_{hom}$ (and it symmetric $\mathsf{BD}_{hom}$), which is a maximal proper fragment of
 $\mathsf{BE}_{hom}$ has been recently proved to be PSPACE-complete
(see \cite{DBLP:conf/icalp/BozzelliMMPS17,DBLP:conf/mfcs/BozzelliMPS20,BMPS21}). 
In this paper, we provide the first known fragment of $\mathsf{HS}_{hom}$ not 
including both the prefix and suffix modalities ($\mathsf{B}$ and $\mathsf{E}$) in the  EXPSPACE-completeness class. 
Such a fragment is the logic $\mathsf{ABD}_{hom}$ which is the extension 
of $\mathsf{BD}_{hom}$ with the \emph{meet} modality $\mathsf{A}$
relating pairs of intervals where one interval begins exactly where the other ends).

In this paper we provide two main novel 
results: (i) we prove that
the satisfiability problem for
$\mathsf{ABD}_{hom}$ is EXPSPACE-hard on finite models
by a reduction to the exponential corridor tiling problem; (ii) we provide a
small model theorem for finite 
models of  $\mathsf{ABD}_{hom}$ formulas
that is doubly exponential in the size of the input formula. Then, by means of such small model theorem, 
we prove that there exists a
decision procedure for the satisfiability problem of $\mathsf{ABD}_{hom}$ formulas
that works using only exponential space w.r.t. the size of the input formula. 

The paper is structured as follows.
In Section~\ref{sec:logic}, we introduce syntax and semantics of $\mathsf{ABD}_{hom}$ under the homogeneity assumption.
In Section \ref{sec:hardness}
we prove that the satisfiability problem
for $\mathsf{ABD}_{hom}$ interpreted over finite models is EXPSPACE-hard.
In Section \ref{sec:compass}, we introduce the notion of homogeneous compass structure, that provides a particularly useful representation for models of $\mathsf{ABD}_{hom}$ formulas. 
In Section~\ref{sec:expspace}, we give an EXPSPACE decision procedure 
for checking the satisfiability of $\mathsf{ABD}_{hom}$ formulas. 
Finally, in Section~\ref{sec:conclusions}, we provide an assessment of the work done and outline future research directions.

\section{The logic $\mathsf{ABD}_{hom}$ }\label{sec:logic}

In this section, we introduce the logic
$\mathsf{ABD}_{hom}$ and we  define the satisfiability relation under the homogeneity assumption.

$\mathsf{ABD}_{hom}$ formulas are built up from a countable  set $\Prop$ of proposition
letters according to the following grammar:
$
\varphi ::= p\ |\ \neg \psi\ |\  \psi \vee \psi \ |\
\hsEA  \psi \ |\ \hsEB \psi \ |\ \hsED \psi,
$
where $p \in \Prop$ and $\hsEA$$,\hsEB$, and $\hsED$ are 
the modalities for Allen's relations \emph{Adjacent} (meets), 
\emph{Begins}, and \emph{During}, respectively.

Let $N \in \bN$ be a natural number and let $\bI_N =\{[x,y]: 0\le x\le y\le N\}$ be the set of all intervals over the 
prefix $0\ldots N$ of $\bN$.
A (finite) model for $\mathsf{BD}$ formulas is a
pair $\bfM = (N, \cV)$, where  $\cV: \bI_N
\rightarrow 2^{\Prop}$ is a valuation that
maps intervals in $\bI_N$ to sets of proposition letters.
Let $\bfM$ be a model and $[x,y]$ an interval.
The semantics of a $\mathsf{ABD}_{hom}$ formula
is defined as follows:
\begin{itemize}
\item $\bfM, [x,y] \models p$ iff $p \in \cV([x,y])$;
\item $\bfM, [x,y] \models \neg \psi$ iff $\bfM,
[x,y] \not\models  \psi$;
\item $\bfM, [x,y] \models \psi_1 \vee \psi_2$ iff
$\bfM, [x,y] \models  \psi_1$ or $\bfM, [x,y] \models  \psi_2$;
\item $\bfM, [x,y] \models \hsEA \psi$ iff 
there is $y'$, with $y'\geq y$, such that
$\bfM, [y,y']\models \psi$;
\item $\bfM, [x,y] \models \hsEB \psi$ iff 
there is $y'$, with $x\leq y'<y$, such that 
$\bfM, [x,y'] \models  \psi$;
\item $\bfM, [x,y] \models \hsED \psi$ iff 
there are $x'$ and $y'$, with $x<x'\leq y'<y$, such that $\bfM, [x',y'] \models  \psi$.
\end{itemize}
The logical constants $\top$ (true) and $\bot$ (false), the Boolean operators $\wedge, \rightarrow$, and $\leftrightarrow$, and the (universal) dual modalities $\hsAA$,
$\hsAB$, and $\hsAD$ can be derived in the standard way.
Moreover, it turns out to be useful to define:
a constant $\pi = \hsAB \bot$ that holds only on intervals of the type 
$[x,x]$ (i.e., points); a global operator $\hsA \psi = \psi\wedge \hsAA \psi
\wedge \hsAB \psi \wedge \hsAB\hsAA \psi $ in order to impose 
constraints on all the intervals in the model. 
We say that a $\mathsf{ABD}_{hom}$ formula
$\varphi$ is (finitely) \emph{satisfiable} 
if and only if there exist
a model $\bfM=(N, \cV)$  and an interval $[x,y]$
such that $\bfM, [0,N] \models \varphi$.
We say that a model $\bfM = (\bI_N, \cV)$
is \emph{homogeneous} if $\cV$ satisfies 
the following property:

\smallskip

$\ \ \ \ \ \ \ \ \ \ \ \forall p\in\Prop ~~ \forall [x,y]\in\bI_N ~~
\Big( p\in\cV([x,y]) ~\Leftrightarrow~ \forall z\in[x,y] ~ p\in\cV([z,z]) \Big).$

\smallskip

%!TEX root = ./main.tex 

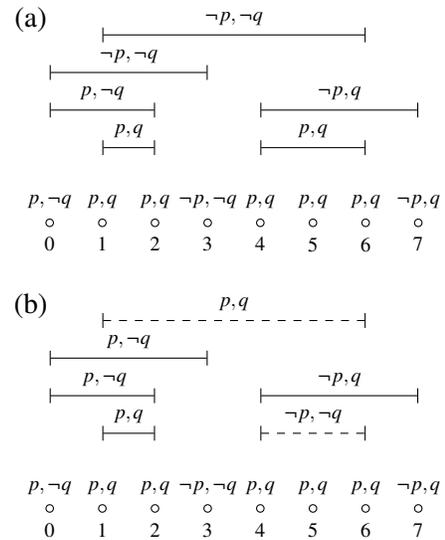
\begin{wrapfigure}{r}{6cm}
    \centering
    \vspace{-0.0cm}
\begin{tikzpicture}
[scale=1,
%background rectangle/.style={fill=red!45}, show background rectangle, 
node distance=0.7cm]

%\newcommand{mathsmall}
% label={[]0:$\scalebox{0.5}{$0$}$}

\node at (-0.25,2.7) {(a)};

\node[draw, circle, inner sep=1, 
label={[]270:$\scalebox{0.7}{$0$}$},
label={[]90:$\scalebox{0.7}{$p,\neg q$}$}](0) {};

\node[draw, circle, inner sep=1, right of=0, 
label={[]270:$\scalebox{0.7}{$1$}$},
label={[]90:$\scalebox{0.7}{$p,q$}$}](1) {};

\node[draw, circle, inner sep=1, right of=1, 
label={[]270:$\scalebox{0.7}{$2$}$},
label={[]90:$\scalebox{0.7}{$p,q$}$}](2) {};

\node[draw, circle, inner sep=1, right of=2, label={[]270:$\scalebox{0.7}{$3$}$},
label={[]90:$\scalebox{0.7}{$\neg p, \neg q$}$}](3) {};

\node[draw, circle, inner sep=1, right of=3, label={[]270:$\scalebox{0.7}{$4$}$},
label={[]90:$\scalebox{0.7}{$p,q$}$}](4) {};

\node[draw, circle, inner sep=1, right of=4, label={[]270:$\scalebox{0.7}{$5$}$},
label={[]90:$\scalebox{0.7}{$p,q$}$}](5) {};

\node[draw, circle, inner sep=1, right of=5, label={[]270:$\scalebox{0.7}{$6$}$},
label={[]90:$\scalebox{0.7}{$p,q$}$}](6) {};

\node[draw, circle, inner sep=1, right of=6, label={[]270:$\scalebox{0.7}{$7$}$},
label={[]90:$\scalebox{0.7}{$\neg p,q$}$}](7) {};

%BEGIN INTERVALS

\draw[|-|] ($(0.center) + (0,2)$) -- ($(3.center)+ (0,2)$)
node[pos=0.5, above](03) {$\scalebox{0.7}{$\neg p, \neg q$}$};

\draw[|-|] ($(0.center) + (0,1.5)$) -- ($(2.center)+ (0,1.5)$)
node[pos=0.5, above](02) {$\scalebox{0.7}{$p, \neg q$}$};

\draw[|-|] ($(1.center) + (0,1)$) -- ($(2.center)+ (0,1)$)
node[pos=0.5, above](12) {$\scalebox{0.7}{$ p, q$}$};

\draw[|-|] ($(4.center) + (0,1)$) -- ($(6.center)+ (0,1)$)
node[pos=0.5, above](46) {$\scalebox{0.7}{$ p, q$}$};

\draw[|-|] ($(4.center) + (0,1.5)$) -- ($(7.center)+ (0,1.5)$)
node[pos=0.5, above](47) {$\scalebox{0.7}{$ \neg p, q$}$};

\draw[|-|] ($(1.center) + (0,2.5)$) -- ($(6.center)+ (0,2.5)$)
node[pos=0.5, above](16) {$\scalebox{0.7}{$ \neg p, \neg q$}$};

%END INTERVALS

%SHIFT POINTS

\pgftransformshift{\pgfpoint{0cm}{-3.8cm}}
\node at (-0.25,2.7) {(b)};
\node[draw, circle, inner sep=1,
label={[]270:$\scalebox{0.7}{$0$}$},
label={[]90:$\scalebox{0.7}{$p,\neg q$}$}](0) {};

\node[draw, circle, inner sep=1, right of=0,
label={[]270:$\scalebox{0.7}{$1$}$},
label={[]90:$\scalebox{0.7}{$p,q$}$}](1) {};

\node[draw, circle, inner sep=1, right of=1,
label={[]270:$\scalebox{0.7}{$2$}$},
label={[]90:$\scalebox{0.7}{$p,q$}$}](2) {};

\node[draw, circle, inner sep=1, right of=2, label={[]270:$\scalebox{0.7}{$3$}$},
label={[]90:$\scalebox{0.7}{$\neg p, \neg q$}$}](3) {};

\node[draw, circle, inner sep=1, right of=3, label={[]270:$\scalebox{0.7}{$4$}$},
label={[]90:$\scalebox{0.7}{$p,q$}$}](4) {};

\node[draw, circle, inner sep=1, right of=4, label={[]270:$\scalebox{0.7}{$5$}$},
label={[]90:$\scalebox{0.7}{$p,q$}$}](5) {};

\node[draw, circle, inner sep=1, right of=5, label={[]270:$\scalebox{0.7}{$6$}$},
label={[]90:$\scalebox{0.7}{$p,q$}$}](6) {};

\node[draw, circle, inner sep=1, right of=6, label={[]270:$\scalebox{0.7}{$7$}$},
label={[]90:$\scalebox{0.7}{$\neg p,q$}$}](7) {};

%BEGIN INTERVALS

\draw[|-|] ($(0.center) + (0,2)$) -- ($(3.center)+ (0,2)$)
node[pos=0.5, above](03) {$\scalebox{0.7}{$p, \neg q$}$};

\draw[|-|] ($(0.center) + (0,1.5)$) -- ($(2.center)+ (0,1.5)$)
node[pos=0.5, above](02) {\scalebox{0.7}{ $p, \neg q$}};

\draw[|-|] ($(1.center) + (0,1)$) -- ($(2.center)+ (0,1)$)
node[pos=0.5, above](12) {$\scalebox{0.7}{$ p, q$}$};

\draw[|-|, dashed] ($(4.center) + (0,1)$) -- ($(6.center)+ (0,1)$)
node[pos=0.5, above](46) {$\scalebox{0.7}{$ \neg p, \neg q$}$};

\draw[|-|] ($(4.center) + (0,1.5)$) -- ($(7.center)+ (0,1.5)$)
node[pos=0.5, above](47) {$\scalebox{0.7}{$ \neg p, q$}$};

\draw[|-|, dashed] ($(1.center) + (0,2.5)$) -- ($(6.center)+ (0,2.5)$)
node[pos=0.5, above](16) {$\scalebox{0.7}{$ p, q$}$};

\end{tikzpicture}

\caption{\label{fig:homvsgen} A homogeneous model (a - above) vs.\ a general one (b - below).}
\vspace{-0.4cm}

\end{wrapfigure}

In Fig.~\ref{fig:homvsgen}, we show a homogeneous model (a) and a  non-homogeneous one (b). 
In homogeneous models, for any proposition letter, the labelling of point-intervals determines that of arbitrary intervals. This is not the case with arbitrary models (see, e.g., $[4,6]$). As a consequence, in homogeneous models, the labelling of the intersection of two intervals contains the labellings of the two intervals (this is the case with intervals $[1,6]$ and $[4,7]$ in Fig.~\ref{fig:homvsgen} (a), whose intersection is the interval $[4,6]$). Once again, this is not the case with arbitrary models (see the very same intervals in Fig.~\ref{fig:homvsgen} (b)).

Satisfiability can be recast in the case of homogeneous models. 
We say that a $\mathsf{ABD}_{hom}$ formula $\varphi$ is \emph{satisfiable under homogeneity} if there 
is a homogeneous model $\bfM$ % = (\bI_N, \cV)$ 
such that $\bfM, [0, N] \models \varphi$.

Satisfiability under homogeneity is clearly more restricted than plain satisfiability. We know from \cite{DBLP:journals/fuin/MarcinkowskiM14,DBLP:conf/icalp/MarcinkowskiMK10} that dropping the homogeneity assumption makes $\mathsf{D}$ undecidable. 
This is not the case with the fragment $\mathsf{B}$
that, being extremely weak in terms of expressive power, remains decidable \cite{DBLP:journals/jancl/GorankoMS04}.
The rest of this paper is devoted to prove the following theorem.

\begin{theorem}\label{thm:expspacecompleteness}
Given a $\mathsf{ABD}_{hom}$ formula $\varphi$ 
the problem of deciding whether or not there exists 
an homogeneous model $\bfM=(N, \cV)$  s.t.
$\bfM, [0,N] \models \varphi$ is an EXPSPACE-complete
problem.
\end{theorem}

The proof of Theorem~\ref{thm:expspacecompleteness} is distributed over
the next three sections. First, in
Section~\ref{sec:hardness}, we prove that such problem is
EXPSPACE-hard,  then, in Section~\ref{sec:compass}
we provide a representation of homogeneous models, called \emph{compass structure}, exploited in Section~\ref{sec:expspace} to design an EXPSPACE decision procedure for the
satisfiability  problem of $\mathsf{ABD}_{hom}$
formulas.

%!TEX root = ./main.tex

\section{EXPSPACE-hardness for the logic
$\mathsf{ABD}_{hom}$ over finite linear orders }\label{sec:hardness}

In this section we prove that the satisfiability
problem for $\mathsf{ABD}_{hom}$ interpreted over
finite linear orders is EXPSPACE-hard. The result
is obtained by a reduction from the \emph{exponential-corridor
tiling problem}, which is known to be EXPSPACE-complete
\cite{van1997convenience}. Such a problem can be stated
as follows.

\begin{problem}\label{prob:exptiling}
Given a tuple $\cT = (T, \tileH,\tileV,C)$ where $T, C \in \bbN$ ($C$ is expressed in binary),
and $\tileH,\tileV\subseteq \{0, \ldots, T\}\times \{0, \ldots, T\}$,
the exponential-corridor
tiling problem consists of
determining whether or not there exists a function
$tile: \bbN \times \{0,\ldots, C\}\rightarrow \{0, \ldots, T\}$ such that:
\begin{compactenum}
\item\label{prob:exptiling:topbot} for every $x \in \bbN$ we have
 $tile(x,0) = 0$ and $tile(x,C)=T$;
\item\label{prob:exptiling:hor} for every $x \in \bbN$
and every $0 \leq y \leq C$ 
we have $(tile(x,y),tile(x+1,y)) \in \tileH$;
\item\label{prob:exptiling:ver} for every $x \in \bbN$
and every $0 \leq y < C$ 
we have $(tile(x,y),tile(x,y+1)) \in \tileV$.
\end{compactenum}
\end{problem}

\vspace{-0.0cm}

\noindent The following classical result 
will be exploited to prove the main goal of this section.

\vspace{-0.0cm}

\begin{theorem}\label{thm:tilingexpspacehard}\cite{van1997convenience}
The exponential-corridor
tiling problem is EXPSPACE-hard.
\end{theorem}

\vspace{-0.0cm}

For defining a reduction 
from Problem~\ref{prob:exptiling} to  the finite satisfiability of
$\mathsf{ABD}_{hom}$ we have to deal with the problem that
the formulas of $\mathsf{ABD}_{hom}$ are interpreted over finite domains
whereas the $tile$ functions ranges over an infinite domain. Roughly
speaking, we shall solve 
Problem~\ref{prob:exptiling} by means of an infinite 
``unfolding'' of  a finite portion of the tiling space that can be  encoded by a (finite) model for a suitable $\mathsf{ABD}_{hom}$ formula. 
The following result is crucial to that purpose.

\vspace{-0.0cm}

\begin{lemma}\label{lem:finitecorridor}
Given an instance $\cT = (T, \tileH,\tileV,C)$ of Problem~\ref{lem:finitecorridor}
we have that $\cT$ is a positive instance if and only if there 
exists a function $tile:\bbN \times \{0,\ldots, C\}\rightarrow 
\{0, \ldots, T\}$ that fulfills conditions \ref{prob:exptiling:topbot},
\ref{prob:exptiling:hor}, and \ref{prob:exptiling:ver} of 
Problem~\ref{prob:exptiling} together with the following one:
\vspace{-0.0cm}
\begin{enumerate}[start={\getrefnumber{prob:exptiling:ver}}]
\addtocounter{enumi}{1}
\item\label{lem:finitecorridor:period} 
there exist $prefix\in \bbN$ and $period\in\bbN^+$ 
s.t. for every $x \geq prefix$ and every $0\leq y\leq C$  we have
$tile(x,y)=tile(x+period,y)$.
\end{enumerate}
\end{lemma}
\vspace{-0.0cm}

The proof of Lemma~\ref{lem:finitecorridor} is straightforward
and omitted. Lemma~\ref{lem:finitecorridor} allows us to 
bound the search space for the existence of the function $tile$
to a finitely representable function $\otile:
\{0,\ldots, prefix, \ldots, prefix + period \} \rightarrow \{0, \ldots, T\}$
for some $prefix \geq 0$ and $period >0$. Function $\otile$
witnesses that $\cT$ is a positive instance of 
Problem~\ref{prob:exptiling} if it satisfies conditions \ref{prob:exptiling:topbot}, \ref{prob:exptiling:hor}, and \ref{prob:exptiling:ver} restricted to
$(x,y)\in \bbN \times \{0, \ldots, C\}$ with $x < prefix + period$
plus the condition that $\otile(prefix, y)=\otile(prefix+period,y)$
for every $y \in\{0, \ldots, C\}$.

Given an instance $\cT = (T, \tileH,\tileV,C)$ of 
Problem~\ref{prob:exptiling} we
provide  a $\textsf{ABD}_{hom}$ formula $\varphi_{\cT}$ 
that is satisfiable over finite models if and  only
if there exists a function $\otile$ that satisfies the
aforementioned properties an thus, by Lemma\ref{lem:finitecorridor},
if and only if $\cT$ is a positive instance of Problem~\ref{prob:exptiling}.
In the proposed encoding we force each point of the model to 
represent exactly one tile. This is done by exploiting $T + 1$
propositional variables $t_0, \ldots, t_T$, called \emph{tile variables}, 
constrained by the following formulas:

\vspace{-0.5cm}

\[\begin{array}{l}
\psi_\exists= \hsA\left(\pi \rightarrow \bigvee\limits_{i=0}^T t_i
\right), \mbox{ given a point in the model \emph{at least} 
one tile variable holds over it;} \\
\arraycellcomment{\psi_!=\hsA \left( \bigwedge\limits_{i=0}^T
\left(t_i \wedge \pi \rightarrow \left(\bigwedge\limits_{j=0, j \neq i}^T
  \neg t_j  \right)
 \right)\right),}{8cm}{given a point in the model \emph{at most}
 one tile variable holds over it (i.e., mutual exclusion).}
 \end{array}\]

\vspace{-0.0cm}

Let us assume w.l.o.g. that $C= 2^c - 1$ for some $c \in \bbN$. Then, we
associate to each model point a number in $\{0, \ldots, C\}$ by 
a binary encoding via $c$-propositional variables
$b_1,\ldots, b_c$, where $b_1$ is the most significative bit. Formally, 
given a model $\bfM = (N, \cV)$ and a point we define a function with

$bit_{\cV}: \{0, \ldots, N\}\times \{b_1,\ldots,b_c \}
\rightarrow \{0,1\}$ where $bit_{\cV} (n, b_i) = \left\{\begin{array}{ll}
1 & \mbox{if $b_i \in \cV([n,n])$}\\
0 & \mbox{otherwise}
\end{array}\right.$. 

For the sake of brevity, we denote with $\oy_n$ the natural number whose $c$-bit 
length binary encoding is $bit_{\cV}(n,b_1)\ldots bit_{\cV}(n,b_c)$. We encode the domain of a general
function $\otile:\{0,\ldots, prefix, \ldots, prefix + period \} \rightarrow \{0, \ldots, T\}$ 
into a finite model $\bfM = (N, \cV)$ by enumerating all the points of the grid
$\{0, \ldots,\allowbreak prefix + suffix \} \times \{0, \ldots, C\}$ along the timepoints
$\{0, \ldots, N\}$ of the model in a lexicographical order. 
The formula  $\psi_{\otile} = \psi_{\exists} \wedge \psi_! \wedge \psi_{boundaries} \wedge \psi_\uparrow$ is used to force such constraint
where $\psi_{boundaries}$ and $\psi_\uparrow$ are formulas defined as follows:

\vspace{-0.3cm}

\[\begin{array}{l}
\arraycellcomment{\psi_{boundaries} =
\hsEB\left(\pi \wedge \bigwedge\limits_{i=1}^c \neg b_i\right)
\wedge \hsAA\left(\bigwedge\limits_{i=1}^c b_i\right), }
{8.5cm}
{ every model $\bfM = (N, \cV)$ for $\psi_{boundaries}$  satisfies $\oy_0=0$ and $\oy_N = C$;}\\
%\end{array}\]
%\[\begin{array}{l}
 \arraycellcomment{
 \psi_{\uparrow} =
 \hsA\left( \hsAB \pi \rightarrow
 \left(\bigwedge\limits_{i=1}^c \hsEB b_i  \wedge \left(
 \hsAA \bot  \vee \bigwedge\limits_{i=1}^c \hsEA (\pi \rightarrow \neg b_i)
 \right)\right) \vee \psi^1_{+}  \right),
 }{4.9cm}{for every  $n \in \{0, \ldots, N\}$
 if $\oy_n=C$ then either $n=N$ or $\oy_{n + 1}= 0$,
 if $\oy_n <N $ then $\oy_{n+1} = \oy_{n} +1$; }\\

 \arraycellcomment{
 \psi^i_{+} = \begin{array}{l}
 (\hsEB b_i \rightarrow \hsEA (\pi \wedge \neg b_i)\wedge 
 \psi^{i+1}_{+})
 \wedge \\
 (\hsEB \neg b_i \rightarrow \hsEA b_i \wedge 
 \psi^{i+1}_{=})\end{array},
 }{8.9cm}{
 formula $\psi^i_{+}$ encodes the bit-wise
 increment for every bit $b_i$ with 
 $i\in \{1, \ldots, c - 1\}$; $\psi^1_{+}$
 is triggered by $\psi_{\uparrow}$
 on every interval $[n, n+1]$ with $\oy_{n} < C$;}\\

 \arraycellcomment{
 \psi^c_{+} = 
 \neg \hsEB b_i \wedge \hsEA b_i,
 }{12.2cm}{
 formula $\psi^c_{+}$ encodes the bit-wise
 increment for the bit $b_c$; it is triggered
 by $\psi^{c-1}_{+}$ on every interval 
 $[n, n+1]$ for which $bit_{\cV}(n,b_i)=1$
 for every $1\leq i < c$; }
 \vspace{-0.0cm}\\
 \mbox{\begin{tabular}{p{15.5cm}}
 let us notice that it does not propagate 
 and it handles overflows by creating a contradiction;
 \end{tabular}
 }\\
%\end{array}\]
%\[\begin{array}{l}
 \arraycellcomment{
 \psi^i_{=} = \neg \pi \wedge  
 \bigwedge\limits_{j=i}^{j=c} \left(
 \hsEB( \pi \wedge b_i) \leftrightarrow \hsEA (\pi \wedge b_i)
 \right),
 }{8.6cm}
 { formula $\psi^i_{=}$ holds over an interval 
 $[n,n']$ if and only if $n<n'$ and
 $bit_{\cV}(n,b_j)=bit_{\cV}(n',b_j)$ for every $i\leq j \leq c$; } 
 \vspace{-0.0cm}
  \end{array}\]
\[\begin{array}{l}
 \mbox{\begin{tabular}{p{15.5cm}}
 Note that if $\psi^1_{=}$ holds over $[n,n']$ then $\oy_n = \oy_{n'}$. Formula $\psi^i_{=}$ is used for guaranteeing the correct bitwise increment 
 in formulas $\psi^i_{+}$, moreover it will be used in the following for correctly identifying tiles which are in the $\tileH$ relation.
 \end{tabular}
 }
 \end{array}\]
%
%\vspace{-0.5cm}
%
%
 \noindent It is worth noticing 
 that any model $\bfM = (N, \cV)$  that satisfies 
 $\psi_{\otile} = \psi_{\exists} \wedge \psi_! \wedge \psi_{boundaries}
 \wedge \psi_\uparrow$  fulfills some properties. First of all,
 the interplay  between $\psi_{boundaries}$ and  $\psi_\uparrow$ 
 guarantees that $N$ is a multiple of $(C+1)$ and thus, 
 for suitably chosen $prefix$ and $suffix$, we can associate each point  
 $(x, y) \in \{0, \ldots,\allowbreak prefix + suffix \} \times 
 \{0, \ldots, C\}$ to a point $n \in \{0, \ldots, N\}$ by means of a 
 bijection $map: \{0, \ldots,\allowbreak prefix + suffix \} \times 
 \{0, \ldots, C\}\rightarrow \{0, \ldots, N\}$ defined as
  $map(x,y)=x\cdot (C+1) 
 + y$ (i.e.,  $map^{-1}(n)= (\lfloor \frac{n}{C + 1} \rfloor, n\ \mbox{\%}\ C  )$
 where \% is the integer remainder operation). Moreover,  
 let us observe that for every element $(x, y)$ in the grid, we have 
 that $x$ is just implicitly encoded in the model by $map(x,y)$ 
 (i.e., $x =\lfloor \frac{map(x,y)}{C + 1} \rfloor$), while 
 $y$ is both implicitly encoded (i.e., $x =\lfloor {map(x,y)}\mbox{\%}\ C$) and explicitly encoded by the the values of variables $b_1 \ldots b_c$
 since it is easy to prove that $\psi_{boundaries}
 \wedge \psi_\uparrow$ forces $y = \oy_{map(x,y)}$.
 Finally, the conjuncts $\psi_{\exists} \wedge \psi_!$ ensure  that 
 each point in $n \in \{0, \ldots, N\}\}$, and thus, by means of $map$,
 any point in the grid, is associated with \emph{exactly} one tile,
 that is the unique tile variable  that belongs to $\cV([n,n])$.
 
For the aforementioned properties, if we consider the function 
 $f$ that maps a function  
 $\otile:\{0, \ldots, M\} \allowbreak \times \{ 0, \ldots, C\} \rightarrow \{0, \ldots, T\}$ in the model $\bfM = (M\cdot(C+1), \cV)$ 
 where for every $(x,y) \in \{0, \ldots, M\}\allowbreak  \times \{ 0, \ldots, C\}$
 we have  $t_i \in \cV([map(x,y), map(x,y)]) $ if and only if $\otile(x,y) = i$ and $\oy_{map(x,y)} = y$, it is easy to prove that $f$ is a bijection between the set of all such $\otile$ function, for every $M \in \bbN^+$,
 and the set of all finite models for $\psi_{\otile}$. 
In summary, the detailed description above shows that any model for $\psi_{\otile}$ is basically a way to represent a generic function $\otile:\{0, \ldots, M\}  \times \{ 0, \ldots, C\} \rightarrow \{0, \ldots, T\}$ and that, viceversa, each of such functions is represented by exactly one model of 
$\psi_{\otile}$. The next step is the encoding of the constraints
of Lemma~\ref{lem:finitecorridor} in $\mathsf{AB}_{hom}$
which allow to check whether there exists a function
$\otile$ that witnesses that $\cT$ is a positive instance.
Such  conditions, restricted to the finite case,
are imposed  by the following formulas:

\vspace{-0.6 cm}

\[\begin{array}{l}
\arraycellcomment{
\psi_{0,C} = \hsA \left(
	\left(\left(\pi \wedge \bigwedge\limits_{i=1}^C \neg b_i \right)\rightarrow t_0 \right)\wedge
	\left(\hspace{-0.1cm}\left(\pi \wedge \bigwedge\limits_{i=1}^C  b_i \right)\rightarrow t_T\hspace{-0.1cm}
  \right)\hspace{-0.1cm}
\right),\hspace{-0.1cm}
}{6.8cm}{ 
formula $\psi_{0,C} $ forces condition~\ref{prob:exptiling:topbot} of Problem~\ref{prob:exptiling}, that is,
the bottom tile of each column is $0$ and the top tile of each column is $T$;
}
\end{array}
\]
\[\begin{array}{l}
\arraycellcomment{
\psi_{\tileH} = \hsA \left(
  \pi \wedge \hsEA \neg \pi \rightarrow 
  \hsEA\left( \psi^{min}_=  \wedge \left(\bigvee\limits_{(i,j)\in \tileH}   (\hsEB t_i  \wedge \hsEA t_j )    \right) \right)
\right),
}{5.1cm}{ 
formula $\psi_{\tileH} $ forces condition~\ref{prob:exptiling:hor} of 
Problem~\ref{prob:exptiling}, that is, each pair of  
}\vspace{-0.0cm}\\ 
\mbox{\begin{tabular}{p{15.5cm}}
grid points of type $(x,y), (x+1,y)$ must be labelled with two tiles that are in the
$\tileH$ relation. This is done by taking for each point $n<N$ the minimal interval
$[n,n']$  with $n <n'$ and  $\oy_n = \oy_{n'}$; then, the $\tileH$ relation is forced 
between the pair of tile variables that hold over $[n,n]$ and $[n',n']$, respectively;
\end{tabular}
} %\\
\end{array}
\]
\[\begin{array}{l}
\arraycellcomment{
\psi^{min}_= = \psi^1_=  \wedge \hsAB \neg \psi^1_=,
}
{12.1cm}
{ formula $\psi^{min}_=$ holds over an interval $[n,n']$  if and only if $n <n'$, $\oy_n = \oy_{n'}$, and
does not exist $n<n''<n'$ such that $\oy_n = \oy_{n''}$. Let us notice that, 
for the constraints}\\ 
\mbox{\begin{tabular}{p{15.5cm}}
imposed by $\psi_{\otile}$ we have that
$n' - n = C + 1$ and thus, 
according to the definition  of $map$,  
we have $map^{-1}(n')= (\lfloor \frac{n}{C + 1} \rfloor + 1, n\ \mbox{\%}\ C) $;
then, $\psi^{min}_=$ holds on all and only those intervals whose endpoints represent horizontally adjacent points of the original grid; \end{tabular}}
%\\
\end{array}
\]
\[\begin{array}{l}
\arraycellcomment{
\psi_{\tileV} = \hsA\left( 
\hsAB \pi \wedge \bigvee\limits_{i=1}^{c} \neg b_i  \rightarrow\bigvee\limits_{(i,j)\in \tileV}   (\hsEB t_i  \wedge \hsEA t_j )
\right),
}{7.5cm}{ formula $\psi_{\tileV} $ forces condition~\ref{prob:exptiling:ver} of Problem~\ref{prob:exptiling}, that is,
each pair of grid points of type $(x,y), (x,y+1)$ must be labelled with two tiles that are
in the $\tileV$ relation. The constraint can be easily imposed 
  }%\\ 
  
  \end{array}
\]
\[\begin{array}{l}
  \mbox{\begin{tabular}{p{15.5cm}}
since the encoding ensures that
vertical consecutive points in the grid corresponds to consecutive points in the model. The constraint 
is triggered on all the intervals of the type  $[n, n+1]$, with the exception of the of the ones with $\oy_n = C$.
The constraint imposes that  unique (thanks to $\psi_\exists\wedge \psi_!$)  pair of  tile variables $(t_i, t_j)$  with 
$(t_i) \in \cV([n,n])$ and $(t_j) \in \cV([n',n'])$ must  satisfy $(i,j) \in \tileV$.
%ones with $\oy_n = C$.
%The constraint imposes that  unique (thanks to $\psi_\exists\wedge \psi_!$)  pair of  tile variables $(t_i, t_j)$  with 
%$(t_i) \in \cV([n,n])$ and $(t_j) \in \cV([n',n'])$ must  satisfy $(i,j) \in \tileV$.
\end{tabular}
} 
 \end{array}
\]
\[\begin{array}{l}
\arraycellcomment{
\psi_{prefix} = 
\begin{array}{c}\hsEB\hsEA \left(p  \wedge  \bigwedge\limits_{i=1}^C  ( \hsEB(\pi \wedge \neg b_i) 
 \wedge \hsEA b_i )  \right) \wedge  \\ \hsA\left( p \wedge \pi  \rightarrow 
 \hsEA\left(\psi^1_= \wedge \hsAA  \neg \psi^1_= \wedge \bigwedge\limits_{i=0}^T (\hsEB t_i \leftrightarrow \hsA t_i)
  \right)\right)
  \end{array},
}{4.8cm}{ formula $\psi_{prefix}$ forces condition~\ref{lem:finitecorridor:period} of Lemma~\ref{lem:finitecorridor}, which 
imposes that there are two distinct columns in the grid which are tiled identically and one of such columns
is the last one.}\\
\mbox{\begin{tabular}{p{15.5cm}}
 This is done by means of a propositional letter $p$. The first conjunct 
of formula $\psi_{prefix}$ imposes that there exists an interval $[n,n']$ in the model 
for which $p\in \cV([n,n'])$, $\oy_n=0$, and $\oy_{n'} = C$ (i.e., $p$ ``covers''
at least one column). Moreover, for the homogeneity assumption, 
we have that $p \in \cV([n'',n''])$  for every $n\leq n''\leq n'$.  The second conjunct imposes that 
for each $p$ labelled points $n$ there must exist a point $n'>n$ with $\oy_n= \oy_{n'}$
(this implicitly implies that $n$ is associated to a grid point which does not belong to the last column).
Moreover, formula $\hsAA  \neg \psi^1_=$ imposes that
 $n'$ must belong to the last column. Finally, it is required  that 
 there exists $0\leq i \leq T$ s.t.  $t_i \in \cV([n,n]) \cap \cV([n',n'])$.
\end{tabular}
}
\end{array}\]
 
\noindent Notice that in the above definitions the use of the $\hsEA$ operator enables us to deal with two key aspects: 
\begin{compactenum} 
\item we can predicate on all the intervals $[n,n']$
for any $n,n' \in \{0, \ldots, N\}$, whereas, by using the $\hsEB$
 operator alone,  we could predicate only on intervals of the form
  $[0, n]$;
 \item we can predicate on the ending point 
 of any current interval $[n,n']$, i.e., the interval $[n',n']$. 
 Such a feature is missing in the logic $\mathsf{BD}_{hom}$
 where we can predicate only on the beginning point of 
 any current interval. For instance, the logic $\mathsf{BD}_{hom}$
 cannot express properties like  $\psi^1_=$
 which checks whether the same  set of propositional letters 
 holds over the two ending points of an interval.
\end{compactenum}
 
Let us define now the formula $\varphi_{\cT}$ as 
$\varphi_{\cT}= \psi_{\otile} \wedge \psi_{0,C} \wedge \psi_{\tileH}\wedge \psi_{\tileV} \wedge \psi_{prefix}$. Since the models 
of  $\psi_{\otile}$ represent all and only the possible finite tiling functions for $\cT$ and 
$\psi_{0,C}$,$\psi_{\tileH}$, $\psi_{\tileV}$, $\psi_{prefix}$
select the subset of such functions/models where 
conditions \ref{prob:exptiling:topbot}, \ref{prob:exptiling:hor}, and
\ref{prob:exptiling:ver},  of Problem~\ref{prob:exptiling} together
with condition \ref{lem:finitecorridor:period} of Lemma~\ref{lem:finitecorridor} are fulfilled we can prove the following result.

\begin{theorem}\label{thm:tilingiffsatisfiable}
Given a instance $\cT = (T, \tileH,\tileV,C)$ of Problem~\ref{prob:exptiling}
we have that $\cT$ is a positive instance if and only if the $\textsf{AB}_{hom}$
formula $\varphi_{\cT}$ is satisfiable over finite linear orders.
\end{theorem}

 It is easy to see that the formula $\varphi_{\cT}$ 
may be generated in LOGSPACE.  It suffices to observe that 
we may define a  multitape Turing Machine that performs the reduction
using just a constant amount of working tapes  
each one holding either $\lceil\log_2 T\rceil$ bits or  $c$ bits. 
Finally, from such an observation together with Theorem~\ref{thm:tilingexpspacehard}
we obtain the main result of this section.

 \begin{theorem}\label{thm:abhomhard}
The satisfiability problem for the logic $\textsf{AB}_{hom}$  over finite linear orders 
is EXPSPACE-hard.
\end{theorem} 
 
We conclude this section by pointing out some interesting facts 
that allow us to better understand 
how the homogeneity assumption affects the satisfiability 
problem of the considered $\textsf{HS}$ fragments.
As a matter of fact the complexity of the satisfiability problem for $\textsf{AB}_{hom}$  over finite linear orders
does not change if we consider its general version $\textsf{AB}$  (i.e., without homogeneity assumption 
\cite{DBLP:journals/tcs/BresolinMMSS14}).
On the contrary, the homogeneity assumption marks a deep difference in the fragment $\textsf{ABD}$. In fact, we shall prove in the next section
that the satisfiabilty problem for $\textsf{ABD}_{hom}$ 
 is decidable in exponential space whereas the problem for $\textsf{ABD}$ is known to be undecidable  
 \cite{DBLP:journals/fuin/MarcinkowskiM14,DBLP:conf/icalp/MarcinkowskiMK10}. 
 As for model checking, the model checking problem over finite Kripke structures for $\textsf{AB}_{hom}$ 
 is proved to be PSPACE-complete  \cite{DBLP:journals/tcs/BozzelliMMPS19}
 while in this work we have already proved that the satisfiability problem over finite  linear orders belongs to an higher complexity class (i.e., EXPSPACE). The tight complexity bound for
 the model checking problem over finite Kripke structures for $\textsf{ABD}_{hom}$ is still open:
 we only know that for its three maximal proper fragments $\textsf{AB}_{hom}$, $\textsf{AD}_{hom}$
 and  $\textsf{BD}_{hom}$ the model checkin problem is PSPACE-complete \cite{DBLP:journals/tcs/BozzelliMMPS19,BMPS21}.

%!TEX root = ./main.tex 

\section{Homogeneous compass structures}
\label{sec:compass}

In this section,
we introduce a spatial representation of homogeneous models, called \emph{homogeneous compass structures}, which will be used to prove that the satisfiabilty problem for $\textsf{ABD}_{hom}$ is decidable in exponential space in Section \ref{sec:expspace}.

Let $\varphi$ be a $\mathsf{BDA}_{hom}$ formula. We define the \emph{closure} of $\varphi$, denoted by $\closure$, as the set of all its sub-formulas and of their negations, plus formulas $\pi$ and $\neg \pi$. 
Moreover, we denote with $\TFA = \{\psi: \hsEA \psi \in \closure\}$ the set of all
the arguments  $\psi$ for  $\hsEA \psi$ formulas in $\closure$. For every 
$\mathsf{BDA}_{hom}$ formula $\varphi$, it holds that $\closure \leq 2|\varphi| + 2$ and $|\TFA| \leq |\closure|/2 -1$. 

\noindent A \emph{$\varphi$-atom} (\emph{atom} for short) is a pair $F_{\alpha}=(F,\alpha)$ where:
\begin{compactenum} 

\item $F$  is a maximal subset of $\closure$ that, for all $\psi \in \closure$, satisfies the following three conditions:
(i) $\psi \in F$ if and only if $\neg \psi \notin F$, 
(ii) if $\psi = \psi_1 \vee \psi_2$, then 
$\psi \in F$ if and only if $\{\psi_1 , \psi_2\}\cap F\neq \emptyset$, and
(iii) if $\pi \in F$ then for every $\hsAA \psi \in F$ we have $\psi \in F$;

\item $\alpha$ is a function $\alpha: \TFA \rightarrow \{ \areq, \asat, \abox\}$ that, for all $\psi\in \TFA$,
satisfies the following four conditions: (i)  if $\alpha(\psi) = \abox$ then $\neg \psi \in F$;
(ii) if $\psi \in F$ then $\alpha(\psi)= \asat$; (iii) if $\pi \in F$ 
and  $\alpha(\psi) = \areq$ then  $\hsEA \psi \in F$ and $\psi \notin F$;
(iv) if $\pi \in F$ 
and  $\alpha(\psi) = \asat$ then  $\psi \in F$.

\end{compactenum}

For the sake of simplicity, from now
on when we refer to $F_\alpha$ as a set, we refer to its first 
component $F$. For instance, when we write 
$\psi \in F_\alpha$,  we mean $\psi \in F$.
An atom $F_\alpha$ is \emph{final} iff for every $\psi \in \TFA$ we have 
$\alpha(\psi) \in\{\asat, \abox\}$. Let $\atoms$ be the set of all
$\varphi$-atoms.  We have that $|\atoms| \leq 2^{|\varphi| + 1}\cdot 2^{|\varphi| - 1} =  2^{2|\varphi|}$, where $|\varphi| =|\closure|/2$.

For all $R \in \{A,B, D\}$, we introduce the functions
$\reqR$, $\obsR$, and $\boxR$, that map each atom $F_\alpha\in\atoms$ 
to the following subsets of $\closure$:
\begin{compactitem}
\item $\reqR(F_\alpha) =\{\psi \in \closure: \hsER \psi \in F\}$; 
\item $\obsR(F_\alpha) =\{\psi \in \closure: \hsER \psi \in \closure, \psi \in F\}$;
\item $\boxR(F_\alpha) =\{\psi \in \closure: \hsAR \psi \in F\}$.
\end{compactitem} 
Note that, for each $F_\alpha \in \atoms$ and each formula 
$\psi$, with $\psi \in \{\psi': \hsEA \psi' \in\closure\}$, either $\psi\in\reqA(F_\alpha)$ or $\neg\psi\in\boxA(F_\alpha)$ and similarly for $B$ and $D$ (it implies that $\boxA(\cdot)$, $\boxB(\cdot)$ and $\boxD(\cdot)$ are not strictly necessary and are introduced only for technical convenience). By exploiting functions above, we define two binary relations $\thenB$ and $\thenD$ over $\atoms$ as follows. For all $F_\alpha, G_\beta \in \atoms$ we write
\begin{compactitem} 
\item $F_\alpha \thenB G_\beta$ iff $\reqB(F_\alpha) = \reqB(G_\beta) \cup 
\obsB(G_\beta)$ and
for every $\psi \in \TFA$ we have $\alpha(\psi) = \beta(\psi)$ if
$\beta(\psi) \in \{\asat, \abox\}$ or $\psi \notin F$;
\item $F_\alpha \thenD G_\beta$ iff $\reqD(F_\alpha) \supseteq \reqD(G_\beta) \cup \obsD(G_\beta)$.
\end{compactitem}

Notice that from the definition of $\thenB$ (resp., $\thenD$), it easily follows that $\boxB(F_\alpha) \subseteq G$ (resp., $\boxD(F_\alpha) \subseteq G$). Notice also that $\thenD$ is transitive (by definition of atom, from $\reqR(F_\alpha) \supseteq \reqR(G_\alpha)$, it immediately follows that $\boxR(F_\alpha) \subseteq \boxR(G_\beta)$), while $\thenB$ is not.

\begin{proposition}\label{lem:atomdeterminacy}
For each pair of atoms $F_\alpha,G_\beta \in \atoms$, we have that 
$F=G$ iff $\reqR(F_\alpha) = \reqR(G_\beta)$ for each
$R \in \{A,B,D\}$, and $F \cap \Prop = G \cap \Prop$. 
\end{proposition}

Given $N\in \bbN$, let $\bG_N = \{(x,y) : 0\le x\le y\le N\}$, given 
a formula $\varphi$, a \emph{$\varphi$-compass structure} (compass structure, when $\varphi$ is clear from the context) is a pair $\cG=(N, \cL)$, where $N \in
\bN$, , and $\cL: \bG_N\rightarrow \atoms$ is a labelling function that satisfies the following properties:
\begin{compactitem}
\item (\emph{initial formula}) $\varphi \in \cL(0,N)$;

\item ($A$-\emph{consistency}) for all $0 \leq  x \leq  y \leq N$,
$\reqA(\cL(x,y)) = \reqA(\cL(y,y))$;

\item ($B$-\emph{consistency}) for all $0 \leq x\leq y < N$, $\cL(x,y+1) \thenB \cL(x,y)$; for all $0 \leq x \leq N$, $\reqB(\cL(x,x)) = \emptyset$;

\item ($D$-\emph{consistency}) for all $0 \leq x < x' \leq y' < y \leq N$, $\cL(x,y) \thenD \cL(x',y')$;

\item ($D$-\emph{fulfilment}) for all $0 \leq x\leq y \leq N$ and all $\psi \in \reqD(\cL(x,y))$,
there exist $x < x' \leq y' < y$ such that  $\psi \in \cL(x',y')$;

\item ($A$-\emph{fulfilment}) for every 
$0 \leq x \leq N$  atom $\cL(x,N)$ is final.
\end{compactitem}

Observe that the definition of $\thenB$ and $B$-consistency guarantee that all the existential requests via the relation $B$ (hereafter $B$-requests) are fulfilled in a compass structure. 

We say that an atom $F \in \atoms$ is \emph{$B$-reflexive} (resp., \emph{$D$-reflexive}) if $F \thenB F$ (resp., $F \thenD F$). If $F$ is not $B$-reflexive (resp., $D$-reflexive), it is \emph{$B$-irreflexive} 
(resp., \emph{$D$-irreflexive}). 

Let $\cG=(N, \cL)$ be a compass structure. We define the function $\cP: \bG_N\rightarrow 2^{\Prop}$ such that $\cP(x,y) = \{ p \in \Prop: p \in \cL(x',x') \mbox{ for all $x\leq x'\leq y$} \}$. We say that a $\varphi$-compass structure $\cG=(\bG_N, \cL)$ is \emph{homogeneous} if for all $(x,y)\in\bG_N$, $\cL(x,y) \cap \Prop = \cP(x,y)$. 
Hereafter, we will often write compass structure for homogeneous $\varphi$-compass structure. 

%!TEX root = ./main.tex 

\begin{wrapfigure}{l}{8.2cm}
\centering
\begin{tikzpicture}

%\newcommand{mathsmall}
% label={[]0:$\scalebox{0.5}{$0$}$}

\clip  (-0.9,-9.2) rectangle (8.8,2.8);

\begin{scope}[node distance=1cm]

\node[draw, circle, inner sep=1, 
label={[]270:$\scalebox{0.7}{$0$}$},
label={[]90:$\scalebox{0.7}{$p,\neg q$}$}](0) {};

\node[draw, circle, inner sep=1, right of=0, 
label={[]270:$\scalebox{0.7}{$1$}$},
label={[]90:$\scalebox{0.7}{$p,q$}$}](1) {};

\node[draw, circle, inner sep=1, right of=1, 
label={[]270:$\scalebox{0.7}{$2$}$},
label={[]90:$\scalebox{0.7}{$p,q$}$}](2) {};

\node[draw, circle, inner sep=1, right of=2, label={[]270:$\scalebox{0.7}{$3$}$},
label={[]90:$\scalebox{0.7}{$\neg p, \neg q$}$}](3) {};

\node[draw, circle, inner sep=1, right of=3, label={[]270:$\scalebox{0.7}{$4$}$},
label={[]90:$\scalebox{0.7}{$p,q$}$}](4) {};

\node[draw, circle, inner sep=1, right of=4, label={[]270:$\scalebox{0.7}{$5$}$},
label={[]90:$\scalebox{0.7}{$p,q$}$}](5) {};

\node[draw, circle, inner sep=1, right of=5, label={[]270:$\scalebox{0.7}{$6$}$},
label={[]90:$\scalebox{0.7}{$p,q$}$}](6) {};

\node[draw, circle, inner sep=1, right of=6, label={[]270:$\scalebox{0.7}{$7$}$},
label={[]90:$\scalebox{0.7}{$\neg p,q$}$}](7) {};

%BEGIN INTERVALS

\draw[|-|] ($(0.center) + (0,2)$) -- ($(3.center)+ (0,2)$)
node[pos=0.5, above](03) {$\scalebox{0.7}{$\neg p, \neg q$}$};

\draw[|-|] ($(0.center) + (0,1.5)$) -- ($(2.center)+ (0,1.5)$)
node[pos=0.5, above](02) {$\scalebox{0.7}{$p, \neg q$}$};

\draw[|-|] ($(1.center) + (0,1)$) -- ($(2.center)+ (0,1)$)
node[pos=0.5, above](12) {$\scalebox{0.7}{$ p, q$}$};

\draw[|-|] ($(4.center) + (0,1)$) -- ($(6.center)+ (0,1)$)
node[pos=0.5, above](46) {$\scalebox{0.7}{$ p, q$}$};

\draw[|-|] ($(4.center) + (0,1.5)$) -- ($(7.center)+ (0,1.5)$)
node[pos=0.5, above](47) {$\scalebox{0.7}{$ \neg p, q$}$};

\draw[|-|] ($(1.center) + (0,2.5)$) -- ($(6.center)+ (0,2.5)$)
node[pos=0.5, above](16) {$\scalebox{0.7}{$ \neg p, \neg q$}$};

%END INTERVALS
\end{scope}

\pgftransformshift{\pgfpoint{0cm}{-9cm}}

\draw[dashed] (0,0) -- (7.5,7.5);
\draw[step=1.0,black, opacity=0.5, very thin,xshift=-1cm,yshift=-1cm] (0.5,0.5)
grid (8.5,8.5);

\draw[->,dashed] (0,0) -- (0,7.5);
\draw[->,dashed] (0,0) -- (7.5,0);

\fill[ black!20, opacity=0.5] (0,0) -- (7.5,7.5) -- (7.5,0);

\begin{scope}[node distance=1.415cm]

\node[draw, circle, inner sep=1, fill=white, 
label={[xshift=0.3cm]30:$\scalebox{0.7}{$p, \neg q, \hsAB\bot$}$}](0) {};

\node[draw, circle, inner sep=1, above right of=0, fill=white, 
label={[xshift=0.1cm]0:$\scalebox{0.7}{$p, q, \hsAB\bot$}$}](1) {};

\node[draw, circle, inner sep=1, above right of=1, fill=white, 
label={[xshift=0.1cm]0:$\scalebox{0.7}{$p, q, \hsAB\bot$}$}](2) {};

\node[draw, circle, inner sep=1, above right of=2, fill=white, 
label={[xshift=0.1cm]0:$\scalebox{0.7}{$\neg p, \neg q, \hsAB\bot$}$}](3) {};

\node[draw, circle, inner sep=1, above right of=3, fill=white, 
label={[xshift=0.1cm]0:$\scalebox{0.7}{$p,  q, \hsAB\bot$}$}](4) {};

\node[draw, circle, inner sep=1, above right of=4, fill=white, 
label={[xshift=0.1cm]0:$\scalebox{0.7}{$p,  q, \hsAB\bot$}$}](5) {};

\node[draw, circle, inner sep=1, above right of=5, fill=white, 
label={[xshift=0.1cm]0:$\scalebox{0.7}{$p,  q, \hsAB\bot$}$}](6) {};

\node[draw, circle, inner sep=1, above right of=6, fill=white, 
label={[xshift=0.3cm, yshift=-0.1cm]93:$\scalebox{0.7}{$
	\begin{array}{l}	
		\neg p,  q,\\ \hsAB\bot
	\end{array}
$}$}](7) {};

\end{scope}

%BEGIN POINTS

\node[draw, circle, inner sep=1, above of=0, fill=white, 
node distance=3cm,
label={[xshift=0.0cm, yshift=0.3cm]0:$\scalebox{0.65}{$
	\begin{array}{l}	
		\neg p, \neg q, \\ \hsEB\top,\\ \hsAB p,\\ \hsAD q
	\end{array}
$}$}](03C)
{};

\draw[->, opacity=0.3] (03) -- (03C);

\node[draw, circle, inner sep=1, above of=4, fill=white, 
node distance=2cm,
label={[xshift=1.5cm]200:$\scalebox{0.7}{$ p, q, \hsEB\top, \hsAB
p, \hsAD q$}$}](46C)
{};

\draw[->, opacity=0.3] (46) edge[bend right] (46C);

\node[draw, circle, inner sep=1, above of=4, fill=white, 
node distance=3cm,
label={[xshift=-0.1cm]0:$\scalebox{0.65}{$
\begin{array}{l}
 \neg p, q, \hsEB\top,\\
  \hsAB p, \hsAD q
\end{array}
$}$}](47C)
{};

\draw[->] (47C) -- (46C) node[pos=0.5,right, xshift=-0.1cm]{\scalebox{0.7}{$B$}};
\draw[->, opacity=0.3] (47) edge[bend left, looseness =0.6] (47C.north);

\node[draw, circle, inner sep=1, above of=1, fill=white, 
node distance=5cm,
label={[xshift=0cm, yshift=0.1cm]270:$\scalebox{0.65}{$ 
\begin{array}{l}
\neg p,\\ \neg q, \\
\hsEB\top,\\ \hsEB\neg p, \\
\hsED\neg q
\end{array}$}
$}](16C)
{};

\draw[->, opacity=0.3] (16) edge[bend left] (16C);

\node[draw, circle, inner sep=1, above of=0, fill=white, 
node distance=2cm,
label={[xshift=0.3cm, yshift=0.05cm]270:$\scalebox{0.65}{$  
	\begin{array}{l}	
		p, \neg q, \hsEB\top, \\
		\hsAB p, \hsAD q
	\end{array}
$}$}](02C)
{};

\draw[->, opacity=0.3] (02) edge[bend right] (02C);

\node[draw, circle, inner sep=1, above of=1, fill=white, 
node distance=1cm,
label={[xshift=-0.2cm, yshift=0.5cm]0:$\scalebox{0.65}{$
	\begin{array}{l}	
		p,  q, \\ \hsEB\top, \\
		\hsAB p, \\ \hsAD q
	\end{array}
$}$}](12C)
{};

\draw[ opacity=0.3] ($(12.south) +(-0.25,0)$) edge[->,bend right, looseness=0.7, out=270, in=270] (12C);

\draw[->] (03C) -- (12C) node[pos=0.5,right, xshift=-0.1cm]{\scalebox{0.7}{$D$}};

%END POINTS

\begin{scope}[opacity=0.5]

\node[yshift=-0.3cm](0) {$\scalebox{0.7}{$0$}$};
\node[ right of=0](0) {$\scalebox{0.7}{$1$}$};
\node[ right of=0](0) {$\scalebox{0.7}{$2$}$};
\node[ right of=0](0) {$\scalebox{0.7}{$3$}$};
\node[ right of=0](0) {$\scalebox{0.7}{$4$}$};
\node[ right of=0](0) {$\scalebox{0.7}{$5$}$};
\node[ right of=0](0) {$\scalebox{0.7}{$6$}$};
\node[ right of=0](0) {$\scalebox{0.7}{$7$}$};
\node[ right of=0, xshift=-0.6cm, yshift=-0.cm, opacity=1](0) {$\scalebox{0.7}{$x$}$};

\node[xshift=-0.3cm](0) {$\scalebox{0.7}{$0$}$};
\node[ above of=0](0) {$\scalebox{0.7}{$1$}$};
\node[ above of=0](0) {$\scalebox{0.7}{$2$}$};
\node[ above of=0](0) {$\scalebox{0.7}{$3$}$};
\node[ above of=0](0) {$\scalebox{0.7}{$4$}$};
\node[ above of=0](0) {$\scalebox{0.7}{$5$}$};
\node[ above of=0](0) {$\scalebox{0.7}{$6$}$};
\node[ above of=0](0) {$\scalebox{0.7}{$7$}$};
\node[ right of=0, xshift=-0.7cm, yshift=0.7cm, opacity=1](0) {$\scalebox{0.7}{$y$}$};

\end{scope}

\draw[line width=0, pattern=north west lines, pattern color=black] (2,5) -- (5,5) -- (2,2);
\draw[->,dashed] (16C) -- (2,5);

\end{tikzpicture}

\caption{\label{fig:modeltocompass} A homogeneous model and the corresponding compass structure. }
\vspace{0.1cm}

\end{wrapfigure}
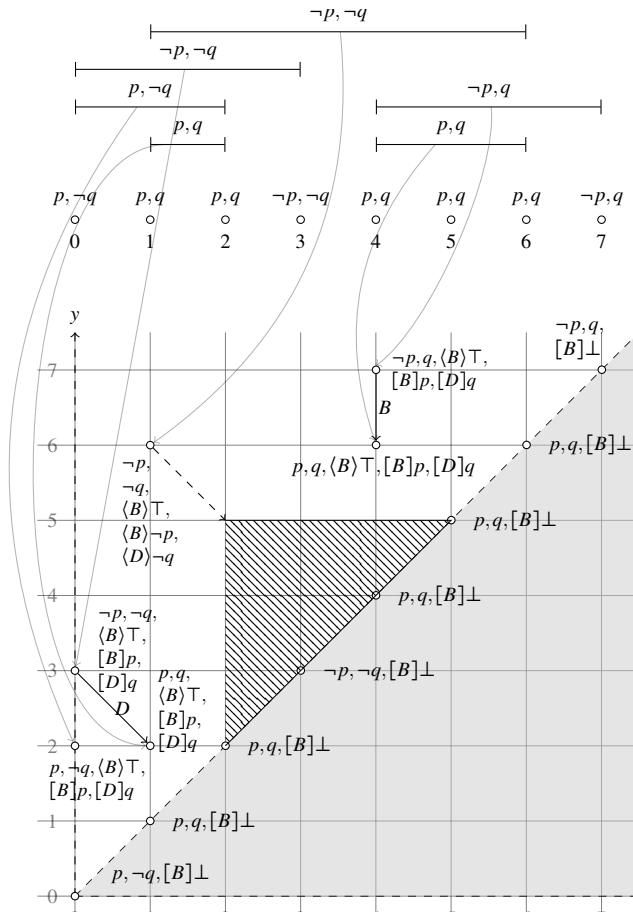

Figure~\ref{fig:modeltocompass} depicts the
homogeneous model $\bfM =(7, \cV)$ of Figure~\ref{fig:homvsgen} $(a)$ with  the corresponding compass structure $\cG=(7, \cL)$, for a given formula $\varphi$.
We assume that $\closure \cap \Prop= \{p,q\}$, $\{\hsEB \psi \in \closure\}= \{ \hsEB \top, \hsEB\neg p \}$, and $\{\hsED \psi \in \closure\}= \{ \hsED \neg q \}$. We know that, by the homogeneity assumption, the valuation of proposition letters at point-intervals determines that at non-point ones.

As an example, if an interval $[x,y]$ contains time point $3$, as, e.g., the interval $[1,6]$, then $\{ p,  q\} \cap\cV([x,y])= \emptyset$. Similarly, if an interval $[x,y]$ contains time point $7$ (resp., $0$), then it must satisfy $\{ p\} \cap\cV([x,y])= \emptyset$ (resp., $\{ q\} \cap\cV([x,y])= \emptyset$). 
As for the compass structure $\cG$, we first observe that each interval $[x,y]$ in $\bfM$ is mapped to a point in the second octant of the $\bN\times\bN$ grid (in Figure~\ref{fig:modeltocompass}, we depict the first quadrant of such a grid, where the first octant is  shaded). 
Analogously, interval relations are mapped into special relations between points (by a slight abuse of terminology, we borrow the names of the interval relations).
As an example, 
%in Figure~\ref{fig:modeltocompass} 
point $(0,2)$ begins $(0,3)$.
 Point $(1,6)$ has points $(2,2), (2,3), (3,3),$ $(2,4), (3,4), (4,4), (2,5), (3,5), (4,5),$ $(5,5)$ as sub-intervals (see the hatched triangle).
In general, all points $(x,x)$ are labelled with irreflexive atoms containing $\hsAB \bot$, while all points $(x,y)$, with $x < y$, are labelled with atoms containing $\hsEB \top$. The variety of atoms is exemplified by the following cases. Atom $\cL(0,3)$ is both $B$-irreflexive and $D$-irreflexive, atom $\cL(4,6)$ is both $B$-reflexive and $D$-reflexive, atom $\cL(4,7)$ is $B$-irreflexive ($\boxB(\cL(4,7))= \{p\}$ and $\neg p \in \cL(4,7)$) and $D$-reflexive ($\boxD(\cL(4,7))= \{q\}$ and $q \in \cL(4,7)$), and atom $\cL(0,2)$ is $B$-reflexive ($\boxB(\cL(0,2))=\{p\}$ and $p \in \cL(0,2)$) and $D$-irreflexive ($\boxD(\cL(0,2))=\{q\}$ and $\neg q \in \cL(0,2)$). Finally, it holds that $\cL(4,7)\thenB\cL(4,6)$ 
($\boxB(\cL(4,7))= \{p,q\}$ and $p, q \in \cL(4,6)$)
and $\cL(3,0)\thenD\cL(1,2)$ ($\boxD(\cL(3,0))= \{q\}$
and $q \in \cL(1,2)$).

The following theorem shows that compass structures are proper tools for solving the satisfiability problem (the proof is straightforward and thus omitted). 

\begin{theorem}\label{thm:satthencompass}
A $\mathsf{ABD}_{hom}$ formula $\varphi$ is satisfiable  iff there is a homogeneous $\varphi$-compass structure.
\end{theorem} 
%!TEX root = ./main.tex 

\section{The satisfiability problem for 
$\mathsf{ABD}_{hom}$ is decidable in EXPSPACE}
\label{sec:expspace}

In this section, we show that the problem of checking whether a $\mathsf{ABD}$ formula $\varphi$ is satisfied by some homogeneous model can be decided in exponential space. 
We first prove that either $\varphi$ is unsatisfiable or it is satisfied by a model of at most doubly-exponential size in $|\varphi|$; then, we show that this model of doubly-exponential size can be guessed in single exponential space.

%we give a detailed account of the proof of the following theorem.
\begin{theorem}\label{thm:bound2}
%Let $\varphi$ be a \LogicABDhom\ formula. The problem of deciding whether or not %it is satisfiable belongs to 
%the complexity class 
%EXPSPACE.
Deciding whether a \LogicABDhom\ formula   $\varphi$ is satisfiable is a problem in EXPACE. 
%the complexity class 
%EXPSPACE.Let be . The problem of deciding whether or not %it is satisfiable belongs to 
%the complexity class 
%EXPSPACE.
\end{theorem}
The proof consists of four main steps whose intuition
will be detailed in the following (due to space bounds complete proofs are omitted).

%!TEX root = ./main.tex 
\vspace{0.25cm}
\noindent \textbf{Step 1:} A finite characterisation of columns and of their relationships.
%\label{subsec:columns}
\vspace{-0.25cm}\\

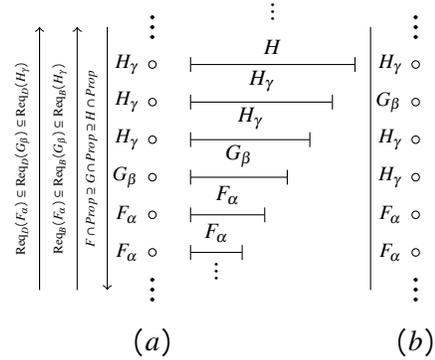
\begin{wrapfigure}{r}{6cm}
    \centering
    \vspace{-0.7cm}
    %!TEX root = ./main.tex 

\begin{tikzpicture}[node distance=0.5cm]

\node[inner sep=0](B) {$\vdots$};

\node[draw, circle, inner sep=1, above of=B, fill=white, 
label={[]180:$\scalebox{0.7}{$F_{\alpha}$}$}](1) {};

\node[draw, circle, inner sep=1, above of=1, fill=white, 
label={[]180:$\scalebox{0.7}{$F_{\alpha}$}$}](2) {};

\node[draw, circle, inner sep=1, above of=2, fill=white, 
label={[]180:$\scalebox{0.7}{$G_{\beta}$}$}](3) {};

\node[draw, circle, inner sep=1, above of=3, fill=white, 
label={[]180:$\scalebox{0.7}{$H_\gamma$}$}](4) {};

\node[draw, circle, inner sep=1, above of=4, fill=white, 
label={[]180:$\scalebox{0.7}{$H_\gamma$}$}](5) {};

\node[draw, circle, inner sep=1, above of=5, fill=white, 
label={[]180:$\scalebox{0.7}{$H_\gamma$}$}](6) {};

\node[inner sep=0, above of=6](T) {$\vdots$};

\draw($(B)+ (-0.6,0)$) edge[<-] node[sloped, above] {$\scalebox{0.45}{$F\cap Prop \supseteq G \cap Prop 
\supseteq H \cap Prop $}$}  ($(T)+ (-0.6,0)$) ; 

\draw($(B)+ (-1,0)$) edge[->] node[sloped, above] {$\scalebox{0.45}{$ \reqB(F_\alpha) \subseteq \reqB(G_\beta) \subseteq \reqB(H_\gamma) $}$}  ($(T)+ (-1,0)$) ; 

\draw($(B)+ (-1.5,0)$) edge[->] node[sloped, above] {$\scalebox{0.45}{$\reqD(F_\alpha) \subseteq \reqD(G_\beta) \subseteq \reqD(H_\gamma)$}$}  ($(T)+ (-1.5,0)$) ;

\draw[|-|] ($(1)+(0.5,0)$) -- ($(1)+(1.2,0)$) node[pos =0.5, above,label={[]270:$\scalebox{0.7}{$\vdots$}$}] {\scalebox{0.75}{$F_{\alpha}$}};  

\draw[|-|] ($(1)+(0.5,0.5)$) -- ($(1)+(1.5,0.5)$) node[pos =0.5, above] {\scalebox{0.75}{$F_{\alpha}$}}; 

\draw[|-|] ($(1)+(0.5,1)$) -- ($(1)+(1.8,1)$) node[pos =0.5, above] {\scalebox{0.75}{$G_{\beta}$}};

\draw[|-|] ($(1)+(0.5,1.5)$) -- ($(1)+(2.1,1.5)$) node[pos =0.5, above] {\scalebox{0.75}{$H_\gamma$}};

\draw[|-|] ($(1)+(0.5,2)$) -- ($(1)+(2.4,2)$) node[pos =0.5, above] {\scalebox{0.75}{$H_\gamma$}};

\draw[|-|] ($(1)+(0.5,2.5)$) -- ($(1)+(2.7,2.5)$) node[pos =0.5, above,label={[]90:$\scalebox{0.7}{$\vdots$}$}] {\scalebox{0.75}{$H$}};

\pgftransformshift{\pgfpoint{0cm}{-0.7cm}}

\node[inner sep=0](C1) {$(a)$};

\pgftransformshift{\pgfpoint{3.5cm}{0cm}}

\node[inner sep=0](C2) {$(b)$};

\pgftransformshift{\pgfpoint{0cm}{0.7cm}}

\draw[thin] (-0.6,0) -- (-0.6,3.5);

\node[inner sep=0](B) {$\vdots$};

\node[draw, circle, inner sep=1, above of=B, fill=white, 
label={[]180:$\scalebox{0.7}{$F_\alpha$}$}](1) {};

\node[draw, circle, inner sep=1, above of=1, fill=white, 
label={[]180:$\scalebox{0.7}{$F_\alpha$}$}](2) {};

\node[draw, circle, inner sep=1, above of=2, fill=white, 
label={[]180:$\scalebox{0.7}{$H_\gamma$}$}](3) {};

\node[draw, circle, inner sep=1, above of=3, fill=white, 
label={[]180:$\scalebox{0.7}{$H_\gamma$}$}](4) {};

\node[draw, circle, inner sep=1, above of=4, fill=white, 
label={[]180:$\scalebox{0.7}{$G_\beta$}$}](5) {};

\node[draw, circle, inner sep=1, above of=5, fill=white, 
label={[]180:$\scalebox{0.7}{$H_\gamma$}$}](6) {};

\node[inner sep=0, above of=6](T) {$\vdots$};

\end{tikzpicture}
    \caption{$(a)$ Monotonicity of atoms along a column in a compass structure, together with a graphical account of the corresponding intervals and of how proposition letters and $B$/$D$ requests 
    must behave. $(b)$ An example of a violation of 
    %the property of 
    monotonicity.
    \vspace{-1cm}
    }
    \label{fig:step1picture}
    \vspace{-1.8cm}
\end{wrapfigure}

\noindent In this section, we first show that, in every compass structure, the atoms that appear in a column $x$ must respect a certain order, that is, they cannot be interleaved. Let $F_\alpha, G_\beta,$ and $H_\gamma$ be three pairwise distinct atoms % $\alpha=\beta =\gamma$ 
with $\reqA(F_\alpha) = \reqA(G_\beta) = H_\gamma$. In Figure \ref{fig:step1picture}.(a), we give a graphical account of the property  to be proved, while, in Figure \ref{fig:step1picture}.(b), we show a violation (atom $H$ appears before and after atom $G$ moving upward along the column).

We preliminarily prove a fundamental property of  $B$-irreflexive atoms. 

\begin{lemma}\label{lem:bstep}
Let $\cG=(N, \cL)$ be a compass structure. For all $x \leq y < N$, if
$\reqB(\cL(x,y))\subset \reqB(\cL(x,y+1))$, then $\cL(x,y)$ is $B$-irreflexive.
\end{lemma}

Given atom $F_\alpha$ in a column, let us now provide a bound on the number
of distinct atoms $G_\beta$ with $\reqA(F_\alpha) = \reqA(G_\beta)$ that 
can be placed above a $F_\alpha$ in a column, that takes into account $B$-requests, $D$-requests, negative literals in $F$, and the number of $\psi \in \TFA$ such that $\alpha(\psi)=\areq$.

\noindent Formally, we define a function $\Deltareq: \atoms \rightarrow \bN$ as follows:
\[
\begin{array}{rcl}
\Deltareq(F_\alpha) &=& (2|\{\hsEB \psi \in \closure \}| - 2|\reqB(F_\alpha)| - \\
&&  |\obsB(F_\alpha) \setminus \reqB(F_\alpha)|)+ \\
&& (|\{\hsED \psi \in \closure \}| - |\reqD(F_\alpha)|) + \\
&& (|\{\neg p: p \in \closure \cap \Prop\}|  - \\
&& |\{ \neg p : p  \in \closure \cap \Prop \wedge  \neg p \in F_\alpha  \}|) + \\
&& |\{\ \psi \in \TFA: \alpha(\psi)=\areq \}|
\end{array}
\]
To understand why a factor $2$ comes into play in the case of $B$-requestes, notice that to move down from an atom including $\hsEB \psi$ to an atom including $\neg \psi, \hsAB \neg \psi$ one must pass through an  atom including $\psi, \hsAB \neg \psi$.
It can be easily checked that, for each $F_\alpha \in \atoms$, $0 \leq \Deltareq(F) \leq 5 |\varphi|$.

To explain how $\Deltareq$ works, we give a simple example. Let $\{\psi: \hsEB \psi \in \closure \} = \{\psi_1\}$ and let $F_\alpha \thenB G_\beta \thenB H_\gamma$,
with $\reqB(H_\gamma)=\{\psi_1\}$ and $\reqB(G_\beta) = \reqB(F_\alpha) = \emptyset$. For simplicity, let  $\TFA = \emptyset$, $\{\psi: \hsED \psi \in \closure \} = \emptyset$, and thus $\reqD(H_\gamma) = \reqD(G_\beta) = \reqD(F_\alpha ) = \emptyset$, and
%(i.e., no $\hsED$ requests in $\varphi$).
$(H_\gamma \cap G_\beta \cap F_\alpha )\cap \Prop= \Prop = \{p\}$. 
It holds that $\Deltareq(F_\alpha )= (2\cdot 1 - 2 \cdot 0 - 0) + (0 - 0) + (1 - 0) + 0= 3$, $\Deltareq(G_\beta)= (2\cdot 1 - 2 \cdot 0 - 1) + (0 - 0) + (1 - 0) + 0= 2$,
and $\Deltareq(H_\gamma)= (2\cdot 1 - 2 \cdot 1 - 0) + (0 - 0) + (1 - 0) + 0= 1$.

We say that an atom $F$ is \emph{initial} if and only 
if $\pi \in F_\alpha$. 
A \emph{$B$-sequence} is a sequence of atoms
$\shadingB= F^0_{\alpha_0}\ldots F^n_{\alpha_n}$ such that:
\begin{inparaenum} 
\item $F^0_{\alpha_0}$ is initial and $F^n_{\alpha_n}$ is final; 
\item for all $0< i \leq n$ we have $F^i_{\alpha_i}\thenB F^{i-1}_{\alpha_{i-1}}$,
$\reqD(F_i) \supseteq \reqD(F_{i-1})$, and
$F_i \cap \Prop \subseteq F_{i-1} \cap \Prop$.
\end{inparaenum}

\noindent A \emph{$B$-sequence} $\shadingB= F^0_{\alpha_0}\ldots F^n_{\alpha_n}$ is \emph{minimal} 
iff for every $0 \leq i < n$ then for every $0\leq i< n$
 $\Deltareq(F^i_{\alpha_i}) > \Deltareq(F^{i+1}_{\alpha_{i+1}})$. Let us observe that 
for every minimal $B$-sequence $\shadingB= F^0_{\alpha_0}\ldots F^n_{\alpha_n}$
we have $n \leq 5|\varphi|$ (i.e., the length of a minimal $B$-sequence is at most 
$5|\varphi|+1$).

\noindent Let $\cG=(N, \cL)$ be a compass structure for $\varphi$ and $0 \leq x \leq N$. We define the
\emph{shading of $x$} in $\cG$, written $\shadingG(x)$, as the sequence of pairs atoms $(\cL(x,y_0), y_0)\ldots(\cL(x,y_m),y_m)$ such that:
\begin{compactenum}
\item $y_i < y_{i+1}$ for every $0\leq i < m$;
\item $\{ \Deltareq(\cL(x,y)): 0 \leq y\leq N\}= \{ \Deltareq(\cL(x,y_i)): 0 \leq i\leq m\}$;
\item for every $0\leq i \leq m$   we have $y_i = \min\left\{ 0 \leq y\leq N: 
  \Deltareq(\cL(x,y_i)) = \Deltareq(\cL(x,y))\right\}$, i.e.,  $y_i$ is the minimum
  height on the column $x$ that exhibits its value for $\Deltareq$.
\end{compactenum} 

\noindent For every  $0 \leq x \leq N$  let $\shadingG(x) = \cL(x,y_0)\ldots\cL(x,y_m)$ 
we denote with $\shadingGB(x)$
the sequence of atoms  $\cL(x,y_0)\ldots\cL(x,y_m)$,
and with  $\shadingGN(x)$  the sequence of natural  numbers $y_0\ldots y_m$, that is, 
the projections of $\shadingG(x)$   of on the first and the second components of its elements, respectively.

\noindent The next lemma easily follows from the definitions of $B$-sequence and shading 
(the proof is omitted).

\begin{lemma}\label{lem:shadingthensequence}
Let $\cG=(N, \cL)$ be a compass structure and 
$0 \leq x \leq N$, then $\shadingGB(x)$
is a minimal $B$-sequence.
\end{lemma} 

%!TEX root = ./main.tex 
\noindent \textbf{Step 2:} Spatial arrangement of atoms between columns. \\

By exploiting the above (finite) characterisation of columns,
we can define a natural equivalence relation of finite index over columns: 
we say that two columns $x, x'$ are equivalent, written $x \sim x'$, if and only if $\shadingGB(x)=\shadingGB(x')$.

\begin{wrapfigure}{l}{6cm}
    \centering
    \vspace{-0.1cm}
    %!TEX root = ./main.tex 

\begin{tikzpicture}[node distance=0.5cm]

\draw[opacity=0.5] (-0.5,-0.5) -- (1.5,1.5);

\node[draw, circle, inner sep=1,  fill=white,
label={[]180:$\scalebox{0.7}{$F^1_{\alpha_1}$}$},
label={[]270:$\scalebox{0.7}{$x$}$}](B) {};

\node[draw, circle, inner sep=1, above of=B, fill=white,
label={[]180:$\scalebox{0.7}{$F^2_{\alpha_2}$}$}](1) {};

\node[draw, circle, inner sep=1, above of=1, fill=white, 
label={[]180:$\scalebox{0.7}{$F^3_{\alpha_3}$}$}](2) {};

\node[draw, circle, inner sep=1, above of=2, fill=white, 
label={[]180:$\scalebox{0.7}{$F^3_{\alpha_3}$}$}](3) {};

\node[draw, circle, inner sep=1, above of=3, fill=white, 
label={[]180:$\scalebox{0.7}{$F^3_{\alpha_3}$}$}](4) {};

\node[draw, circle, inner sep=1, above of=4, fill=white, 
label={[]180:$\scalebox{0.7}{$F^4_{\alpha_4}$}$}](5) {};

\node[draw, circle, inner sep=1, above of=5, fill=white, 
label={[]180:$\scalebox{0.7}{$F^4_{\alpha_4}$}$}](6) {};

\node[inner sep=0, above of=6](T) {$\vdots$};

\pgftransformshift{\pgfpoint{1cm}{1cm}}

\node[draw, circle, inner sep=1,  fill=white,
label={[]0:$\scalebox{0.7}{$F^1_{\alpha_1}$}$},
label={[]270:$\scalebox{0.7}{$x'$}$}](RB) {};

\node[draw, circle, inner sep=1, above of=RB, fill=white,
label={[]0:$\scalebox{0.7}{$F^2_{\alpha_2}$}$}](R1) {};

\node[draw, circle, inner sep=1, above of=R1, fill=white, 
label={[]0:$\scalebox{0.7}{$F^3_{\alpha_3}$}$}](R2) {};

\node[draw, circle, inner sep=1, above of=R2, fill=white, 
label={[]0:$\scalebox{0.7}{$F^3_{\alpha_3}$}$}](R3) {};

\node[draw, circle, inner sep=1, above of=R3, fill=white, 
label={[]0:$\scalebox{0.7}{$F^4_{\alpha_4}$}$}](R4) {};

\node[inner sep=0, above of=R4](T) {$\vdots$};

\draw[opacity=0.2] ($(2) + (-0.25,0)$) -- (2) -- (RB);
\draw[opacity=0.2] ($(3) + (-0.25,0)$) -- (3) -- (R1) -- ($(R1) + (0.25,0)$);
\draw[opacity=0.2] ($(4) + (-0.25,0)$) -- (4) -- (R2) -- ($(R2) + (0.25,0)$);
\draw[opacity=0.2] ($(5) + (-0.25,0)$) -- (5) -- (R3) -- ($(R3) + (0.25,0)$);
\draw[opacity=0.2] ($(6) + (-0.25,0)$) -- (6) -- (R4) -- ($(R4) + (0.25,0)$);

\pgftransformshift{\pgfpoint{-0.3cm}{-1.5cm}}

\node[inner sep=0](C1) {$(a)$};

\pgftransformshift{\pgfpoint{2.5cm}{0cm}}

\node[inner sep=0](C2) {$(b)$};

\pgftransformshift{\pgfpoint{-0.5cm}{0.5cm}}
\draw[thin] (-0.8,-0.5) -- (-0.8,3);

\draw[opacity=0.5] (-0.5,-0.5) -- (1.5,1.5);

\node[draw, circle, inner sep=1,  fill=white,
label={[]180:$\scalebox{0.7}{$F^1_{\alpha_1}$}$},
label={[]270:$\scalebox{0.7}{$x$}$}](B) {};

\node[draw, circle, inner sep=1, above of=B, fill=white,
label={[]180:$\scalebox{0.7}{$F^2_{\alpha_2}$}$}](1) {};

\node[draw, circle, inner sep=1, above of=1, fill=white, 
label={[]180:$\scalebox{0.7}{$F^3_{\alpha_3}$}$}](2) {};

\node[draw, circle, inner sep=1, above of=2, fill=white, 
label={[]180:$\scalebox{0.7}{$F^3_{\alpha_3}$}$}](3) {};

\node[draw, circle, inner sep=1, above of=3, fill=white, 
label={[]180:$\scalebox{0.7}{$F^3_{\alpha_3}$}$}](4) {};

\node[draw, circle, inner sep=1, above of=4, fill=white, 
label={[]180:$\scalebox{0.7}{$F^3_{\alpha_3}$}$}](5) {};

\node[draw, circle, inner sep=1, above of=5, fill=white, 
label={[]180:$\scalebox{0.7}{$F^4_{\alpha_4}$}$}](6) {};

\node[inner sep=0, above of=6](T) {$\vdots$};

\pgftransformshift{\pgfpoint{1cm}{1cm}}

\node[draw, circle, inner sep=1,  fill=white,
label={[]0:$\scalebox{0.7}{$F^1_{\alpha_1}$}$},
label={[]270:$\scalebox{0.7}{$x'$}$}](RB) {};

\node[draw, circle, inner sep=1, above of=RB, fill=white,
label={[]0:$\scalebox{0.7}{$F^2_{\alpha_2}$}$}](R1) {};

\node[draw, circle, inner sep=1, above of=R1, fill=white, 
label={[]0:$\scalebox{0.7}{$F^3_{\alpha_3}$}$}](R2) {};

\node[draw, circle, inner sep=1, above of=R2, fill=white, 
label={[]0:$\scalebox{0.7}{$F^4_{\alpha_4}$}$}](R3) {};

\node[draw, circle, inner sep=1, above of=R3, fill=white, 
label={[]0:$\scalebox{0.7}{$F^4_{\alpha_4}$}$}](R4) {};

\node[inner sep=0, above of=R4](T) {$\vdots$};

\draw[opacity=0.2] ($(2) + (-0.25,0)$) -- (2) -- (RB);
\draw[opacity=0.2] ($(3) + (-0.25,0)$) -- (3) -- (R1) -- ($(R1) + (0.25,0)$);
\draw[opacity=0.2] ($(4) + (-0.25,0)$) -- (4) -- (R2) -- ($(R2) + (0.25,0)$);
\draw[opacity=0.2] ($(5) + (-0.25,0)$) -- (5) -- (R3) -- ($(R3) + (0.25,0)$);
\draw[opacity=0.2] ($(6) + (-0.25,0)$) -- (6) -- (R4) -- ($(R4) + (0.25,0)$);

\node[fit=(5)(R3), draw] {};

\end{tikzpicture}
    \vspace{-0.1cm}
    \caption{Two equivalent columns that respect the order $(a)$ and  two equivalent columns that violates it $(b)$.
    \vspace{-0.5cm}}
    \label{fig:step2picture}
\end{wrapfigure}
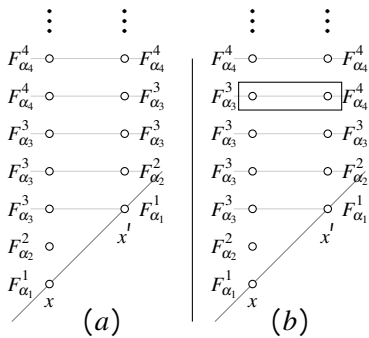
In the following, we prove that equivalent columns can be totally ordered according to a 
given partial order relation over their shadings. 
Formally, for any two equivalent columns $x\sim x'$ let 
$\shadingG(x) = (\cL(x,y_0), y_0)\ldots(\cL(x,y_m),y_m)$ and
$\shadingG(x') = (\cL(x,\oy_0), \oy_0)\ldots(\cL(x,\oy_m),\oy_m)$,  
$\shadingG(x) \leq \shadingG(x')$ if and only if for every 
$0 \leq i \leq m$ we have $y_i \leq \oy_i$.
Intuitively, if we recall that, by definition,  for every $0\leq i\leq m$ the  row $y_i$ (resp. $\oy_i$) is the 
the minimum row for which  atom $\cL(x,y_i)=\cL(x',\oy_i)$ (from $\shadingGB(x)=\shadingGB(x')$)  
occurs on column $x$ (resp. $x'$) meaning that, moving upward  column $x'$, an atom cannot appear until it has appeared on column $x$. In Fig. \ref{fig:step2picture}.(a), we depict two equivalent columns that satisfy such a property.
In general, when moving upward, atoms on $x'$ are often ‘‘delayed'' with respect to atoms in $x$, the limit case being when atoms on the same row are equal.
In Fig. \ref{fig:step2picture}.(b), a violation of the property
(boxed atoms) is shown. 
The following lemma shows that such a violation never occurs in a compass structure.

\begin{lemma}\label{lem:shadingorder}
Let $\cG=(N, \cL)$ be a compass structure.  
For every pair of equivalent columns $x \sim x'$ with $0\leq x<x'\leq N$, it holds that $\shadingG(x) < \shadingG(x')$.
\end{lemma}

%!TEX root = ./main.tex 

\noindent \textbf{Step 3:} $B$-sequence suffixes starting at the same row have  bounded variability in $|\varphi|$.\\

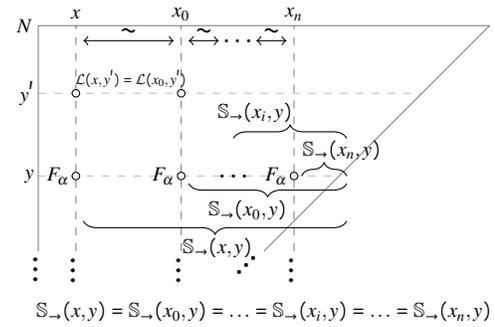
\begin{wrapfigure}{r}{6.5cm}
    \centering
    \vspace{-0.5cm}
    %!TEX root = ./main.tex 

\begin{tikzpicture}[node distance=0.5cm]

\draw[opacity=0.5] (3,3) -- (6,6) node[sloped, pos=-0.07, opacity=1] {$\cdots$};
\draw[opacity=0.25, dashed] (0,4) -- (4,4) node[opacity=1, pos = -0.03] {\scalebox{0.7}{$y$}};
\draw[opacity=0.25, dashed] (0,5.1) -- (5.1,5.1) node[opacity=1, pos = -0.03] {\scalebox{0.7}{$y'$}};
\draw[opacity=0.5] (0,3) -- (0,6) node[sloped, pos=-0.08, opacity=1] {$\cdots$};
\draw[opacity=0.5] (0,6) -- (6,6) node[opacity=1, pos = -0.03] {\scalebox{0.7}{$N$}};

\draw[opacity=0.5, dashed] (0.5,3) -- (0.5,6) node[opacity=1, pos = 1.05] {\scalebox{0.7}{$x$}} node[sloped, pos=-0.085, opacity=1] {$\cdots$};

\draw[opacity=0.5, dashed] (1.9,3) -- (1.9,6) node[opacity=1, pos = 1.05] {\scalebox{0.7}{$x_0$}} node[sloped, pos=-0.085, opacity=1] {$\cdots$};

\draw[opacity=0.5, dashed] (3.4,3) -- (3.4,6) node[opacity=1, pos = 1.05] {\scalebox{0.7}{$x_n$}} node[sloped, pos=-0.085, opacity=1] {$\cdots$};

\draw[<->] (0.6,5.8) -- (1.8, 5.8) node[pos=0.5, above=-0.1cm] {$\sim$};
\draw[<->] (2,5.8) -- (2.4, 5.8) node[pos=0.5, above=-0.1cm,
label={[yshift=-0.1cm, xshift=-0.15cm]0:$\scalebox{1}{$\ldots$}$}] {$\sim$};
\draw[<->] (2.9,5.8) -- (3.3, 5.8) node[pos=0.5, above=-0.1cm] {$\sim$};

\pgftransformshift{\pgfpoint{0.5cm}{4cm}}

%\node[draw, circle, inner sep=1,  fill=white,
%label={[]180:$\scalebox{0.7}{$F_\alpha_1$}$},
%label={[]270:$\scalebox{0.7}{$x$}$}](B) {};

\node[draw, circle, inner sep=1,  fill=white,
label={[label distance=-0.1cm]180:$\scalebox{0.7}{$F_\alpha$}$}](X) {};

\node[draw, circle, inner sep=1,  fill=white,
label={[label distance=0.3cm, name= XI]0:$\scalebox{1}{$\ldots$}$},
label={[label distance=-0.1cm]180:$\scalebox{0.7}{$F_\alpha$}$}, right of=X, node distance=1.4cm](X0) {};

\node[draw, circle, inner sep=1,  fill=white,
label={[label distance=-0.1cm, xshift=-0.1cm, yshift=0.2cm]0:$\scalebox{0.5}{$\cL(x, y')= \cL(x_0, y')$}$}, above of=X, node distance=1.1cm](AX) {};

\node[draw, circle, inner sep=1,  fill=white, above of=X0, node distance=1.1cm](AX) {};

\node[draw, circle, inner sep=1,  fill=white,
label={[label distance=-0.1cm]180:$\scalebox{0.7}{$F_\alpha$}$}, right of=X0, node distance=1.5cm](XN) {};

\draw [decorate,decoration={brace,amplitude=5pt}]
($(XN)+ (0.1,0)$) -- (3.6,0) 
node [black,midway,yshift=0.35cm, xshift =0.2cm](HN) {\scalebox{0.7}{
$\future(x_n, y)$}};

\draw [decorate,decoration={brace,amplitude=5pt}]
($(XI)+ (0,0.5)$) -- (3.6,0.5) 
node [black,midway,yshift=0.35cm, xshift=-0.5cm](HN) {\scalebox{0.7}{
$\future(x_i, y)$}};

\draw [decorate,decoration={brace,amplitude=5pt, mirror}]
($(X0)+ (0.1,-0.1)$) -- (3.6,-0.1) 
node [black,midway,yshift=-0.35cm, xshift=-0.3cm](HN) {\scalebox{0.7}{
$\future(x_0, y)$}};

\draw [decorate,decoration={brace,amplitude=5pt, mirror}]
($(X)+ (0.1,-0.6)$) -- (3.6,-0.6) 
node [black,midway,yshift=-0.35cm, xshift=0cm](HN) {\scalebox{0.7}{
$\future(x, y)$}};
\pgftransformshift{\pgfpoint{2.5cm}{-1.8cm}}

\node {\scalebox{0.7}{$\future(x, y) = \future(x_0, y)= \ldots = \future(x_i, y)= \ldots = \future(x_n, y)$}};

\end{tikzpicture}
    \vspace{-0.45cm}
    \caption{A graphical account of the behaviour of covered points. We have that $x$ is covered by $x_0 < \ldots < x_n$ on row $y$ and thus the labelling of points on column $x$ above $(x,y)$ is exactly the same of the correspondent 
    points on column $x_0$ above $(x_0,y)$, that is, $\cL(x,y')=\cL(x_0,y')$, for all $y\leq y'\leq N$.}
    \label{fig:step3picture}
    \vspace{-0.0cm}
\end{wrapfigure}

\noindent Let us now provide a very strong characterization of the rows in a compass structure by making use of a covering property, depicted in Fig.\ \ref{fig:step3picture}.

\noindent Let $\cG=(N, \cL)$ be a compass structure and let $0\leq x \leq y$. We define $\future(x,y)$ as the set $\{ (\shadingGB(x'), \cL( x',y)): x'>x \}$. %Basically, 
$\future(x,y)$ collects the equivalence classes of $\sim$ which are witnessed to the right of $x$ on row $y$ by means of their 
B-sequence plus a ``pointer'' to the ``current atom'', that
is, the atoms they are exposing on $y$. 
If $\cG=(N, \cL)$ is homogeneous (as in our setting), for all $0 \leq x \leq y\leq N$, the number of possible sets 
$\future(x,y)$ is bounded by 
$2^{6^{5|\varphi|^2 + 2 |\varphi|} \cdot \frac{2}{3}^{5|\varphi| + 2}}$, 
that is, it is doubly exponential in the size of $|\varphi|$. 
For every  $0 \leq x \leq y\leq N$ let us define the \emph{fingerprint of $(x,y)$ in $\cG$}, written
$\fpG(x,y)$, as the triple $\fpG(x,y) = (\shadingGB(x), \cL(x,y), \future(x,y))$.
Lemma \ref{lem:stepconstraint} constrains the way in which two columns $x, x'$, with $x<x'$ and $x \sim x'$,
evolve from a given row $y$ on when  $\fpG(x,y) = \fpG(x',y)$.

\noindent For two atoms $F_\alpha$ and $G_\beta$, we say that 
they are \emph{equivalent modulo 
$\textsf{A}$}, written $F_\alpha \equivmodA G_\beta$ 
if and only if $F \setminus \reqA(F_\alpha) = G \setminus \reqA(G_\beta)$
and $\alpha= \beta$ (i.e., $F_\alpha$ and $G_\beta$
have at most different $\hsEA$ requests).

\begin{lemma}\label{lem:stepconstraint}
Let $\cG=(N, \cL)$ be a compass structure and let
$0\leq x < x'\leq y \leq N $. If $\fpG(x,y) = \fpG(x',y)$
and $y'$ is the smallest point greater than $y$ such that
$\cL(x,y')\nequivmodA \cL(x,y)$, if any, and $N$ otherwise, 
then, for all $y \leq y''\leq y'$, $\cL(x,y'') = \cL(x',y'')$.
\end{lemma}

\noindent From Lemma~\ref{lem:stepconstraint}, the next corollary follows.

\begin{corollary}\label{cor:inbetweeners}
Let $\cG=(N, \cL)$ be a compass structure and let
$0\leq x < x'\leq y \leq N $. If $\fpG(x,y) = \fpG(x',y)$ and
$y'$ is the smallest point greater than $y$ such that
$\cL(x,y')\nequivmodA \cL(x,y)$, if any, and $N$ otherwise, 
then, for every pair of points $\ox, \ox'$, with
$x < \ox < x' < \ox'$, with $\cL(\ox,y)= \cL(\ox', y)$ and
$\ox\sim \ox' \not\sim x$, 
it holds that $\cL(\ox, y'')= \cL(\ox', y'')$, for all $y \leq y'' \leq y'$.
\end{corollary}

The above results lead us to the identification of those points $(x,y)$ 
whose behaviour perfectly reproduces that of a number of points $(x',y)$
on their right with $\fpG(x,y) = \fpG(x',y)$.
These points $(x,y)$, like all points ``above'' them, are irrelevant
with respect to fulfilment in a compass structure. We call them 
\emph{covered points}.

\begin{definition}\label{def:coveredpoint}
 Let $\cG=(N, \cL)$ be a compass structure  
 and $0\leq x\leq y\leq N$. We say that $(x,y)$
 is \emph{covered} iff there exist
 $n+1 = \Deltareq(\cL(x,y))$ distinct points $x_0 < 
 \ldots < x_n \leq y$, with  $x < x_0$, such that for all
$0\leq i \leq n$, $\fpG(x,y) = \fpG(x_i,y)$.
In such a case, we say that $x$ is covered by $x_0 < \ldots < x_n$ on $y$.
\end{definition}

\begin{lemma}\label{lem:coveredstability}
  Let $\cG=(N, \cL)$ be a compass structure  and let
  $x,y$, with $0 \leq x \leq y \leq N$, be two points  
  such that $x$ is covered by points $x_0 < \ldots < x_n$ on $y$. 
  Then, for all $y \leq y'  \leq N$, it holds that  $\cL(x,y') = 
  \cL(x_0,y')$. 
\end{lemma}

%!TEX root = ./main.tex 

\begin{wrapfigure}{l}{8cm}
\centering
\begin{tikzpicture}[scale=0.8]

\clip  (-1.2,-1.2) rectangle (8.8,5.2);

%\draw[dashed] (0,0) -- (8.5,8.5);

\draw[step=0.9,black, opacity=0.5, very thin] (-1.2,-1.2)
grid (13.2,5.2);

\draw[->,dashed] (0,-1.4) -- (0,5.5);

\draw[->,dashed] (6.5,-1.5) -- (13.5,5.5);

\fill[ black!20, opacity=0.5] (6.5,-1.5) -- (13.5,5.5) -- (13.5,-1.5);

\draw[dashed]  (-0.2, 0) -- (10,0) node[pos=0, xshift=-0.2cm] {$y$};
\draw[dashed]  (-0.2, 4.5) -- (14,4.5) node[pos=0, xshift=-0.2cm] {$y'$};
\draw[dashed]  (-0.2, 3.6) -- (14,3.6) node[pos=0, xshift=-0.3cm] {$y' - 1$};

%Y

\node[draw, circle, inner sep=2, 
label={[]270:$\scalebox{0.7}{$x$}$},
label={[xshift=-0.2cm]90:$\scalebox{0.7}{$F^i_{\alpha_i}$}$}, fill=white](X) at (0.9,0) {};

\node[draw, circle, inner sep=2, 
label={[]270:$\scalebox{0.7}{$\ox$}$},
label={[xshift=-0.3cm]90:$\scalebox{0.7}{$G^j_{\beta_j}$}$}, fill=white](OX) at (1.8,0)
{};

\node[draw, circle, inner sep=2, 
label={[]270:$\scalebox{0.7}{$x_0$}$},
label={[xshift=-0.2cm]90:$\scalebox{0.7}{$F^i_{\alpha_i}$}$}, fill=white](X1) at (2.7,0)
{};

\node[](D) at (3.6,0)
{$\scalebox{2}{$\ldots$}$};

\node[draw, circle, inner sep=2, 
label={[]270:$\scalebox{0.7}{$x_{n-1}$}$},
label={[xshift=-0.3cm]90:$\scalebox{0.7}{$F^i_{\alpha_i}$}$}, fill=white](XNN) at (4.5,0)
{};

\node[inner sep=0](XNND) at (4.5,2) {$\scalebox{2}{$\vdots$}$};

\node[draw, circle, inner sep=2, 
label={[]270:$\scalebox{0.7}{$ $}$},
label={[xshift=-0.3cm]90:$\scalebox{0.7}{$F^{i+1}_{\alpha_{i+1}}$}$}, fill=white](XNN1) at (4.5,4.5)
{};

\draw[dashed] (XNN1) -- ($(XNND)+(0,0.1)$);
\draw[dashed] (XNND) -- (XNN);

\node[draw, circle, inner sep=2, 
label={[]270:$\scalebox{0.7}{$ $}$},
label={[xshift=-0.3cm]90:$\scalebox{0.7}{$F^i_{\alpha_i}$}$}, fill=white](XNN2) at (4.5,3.6)
{};

\node[draw, circle, inner sep=2, 
label={[]270:$\scalebox{0.7}{$x_n$}$},
label={[xshift=-0.3cm]90:$\scalebox{0.7}{$F^i_{\alpha_i}$}$}, fill=white](XN) at (5.4,0)
{};

\node[draw, circle, inner sep=2, 
label={[]270:$\scalebox{0.7}{$\ox'$}$},
label={[xshift=-0.3cm]90:$\scalebox{0.7}{$G^j_{\beta_j}$}$}, fill=white](OXP) at (6.3,0)
{};

\node[draw, circle, inner sep=2, 
label={[]270:$\scalebox{0.7}{$\hx$}$},
label={[xshift=-0.3cm]90:$\scalebox{0.7}{$F^i_{\alpha_i}$}$}, fill=white](HX) at (7.2,0)
{};

%Y'-1

\node[draw, circle, inner sep=2, 
label={[]270:$\scalebox{0.7}{}$},
label={[xshift=-0.2cm]90:$\scalebox{0.7}{$F^i_{\alpha_i}$}$}, fill=white](XYP1) at (0.9,3.6)
{};

\node[inner sep=0](DX) at (0.9,2) {$\scalebox{2}{$\vdots$}$};

\node[draw, circle, inner sep=2, 
label={[]270:$\scalebox{0.7}{$ $}$},
label={[xshift=-0.3cm]90:$\scalebox{0.7}{$G^{j+k}_{\beta_{j+k}}$}$}, fill=white](OXYP1) at (1.8,3.6)
{};

\draw[dashed] (XYP1) -- ($(DX)+(0,0.1)$);
\draw[dashed] (DX) -- (X);

\node[draw, circle, inner sep=2, 
label={[]270:$\scalebox{0.7}{$ $}$},
label={[xshift=-0.3cm, yshift=-0.1cm]90:$\scalebox{0.7}{$G^{j+k-1}_{\beta_{j+k-1}}$}$}, fill=white](OXYPM1) at
(1.8,2.7)
{};

\node[draw, circle, inner sep=2, 
label={[]270:$\scalebox{0.7}{$ $}$},
label={[xshift=-0.3cm]90:$\scalebox{0.7}{$G^{j+1}_{\beta_{j+1}}$}$}, fill=white](OXYP1)
at
(1.8,0.9)
{};

\node[inner sep=0](DX) at (1.8,2) {$\scalebox{2}{$\vdots$}$};

\node[draw, circle, inner sep=2, 
label={[]270:$\scalebox{0.7}{}$},
label={[xshift=-0.2cm]90:$\scalebox{0.7}{$F^i_{\alpha_i}$}$}, fill=white](X1YP1) at (2.7,3.6)
{};

\node[inner sep=0](DX1) at (2.7,2) {$\scalebox{2}{$\vdots$}$};

\draw[dashed] (X1YP1) -- ($(DX1)+(0,0.1)$);
\draw[dashed] (DX1) -- (X1);

\node[](D) at (3.6,3.6)
{$\scalebox{2}{$\ldots$}$};

\node[](D) at (3.6,4.5)
{$\scalebox{2}{$\ldots$}$};

\node[draw, circle, inner sep=2, 
label={[]270:$\scalebox{0.7}{}$},
label={[xshift=-0.2cm]90:$\scalebox{0.7}{$F^i_{\alpha_i}$}$}, fill=white](XNYP1) at
(5.4,3.6)
{};

\node[inner sep=0](DXN) at (5.4,2) {$\scalebox{2}{$\vdots$}$};

\draw[dashed] (XNYP1) -- ($(DXN)+(0,0.1)$);
\draw[dashed] (DXN) -- (XN);

\node[draw, circle, inner sep=2, 
label={[]270:$\scalebox{0.7}{$ $}$},
label={[xshift=-0.3cm]90:$\scalebox{0.7}{$G^{j+1}_{\beta_{j+1}}$}$}, fill=white](OXP1)
at (6.3,0.9)
{};

\node[draw, circle, inner sep=2, 
label={[]270:$\scalebox{0.7}{$ $}$},
label={[xshift=-0.3cm]90:$\scalebox{0.7}{$G^{j+1}_{\beta_{j+1}}$}$}, fill=white](OXP1)
at (6.3,0.9)
{};

\node[inner sep=0](DOXP) at (6.3,2) {$\scalebox{2}{$\vdots$}$};

\node[draw, circle, inner sep=2, 
label={[]270:$\scalebox{0.7}{$ $}$},
label={[xshift=-0.3cm, yshift=-0.1cm]90:$\scalebox{0.7}{$G^{j+k-1}_{\beta_{j+k-1}}$}$}, fill=white](OXYPM1)
at (6.3,2.7)
{};

\node[draw, circle, inner sep=2, 
label={[]270:$\scalebox{0.7}{$ $}$},
label={[xshift=-0.3cm]90:$\scalebox{0.7}{$G^{j+k}_{\beta_{j+k}}$}$}, fill=white](OXYP1)
at (6.3,3.6)
{};

\node[draw, circle, inner sep=2, 
label={[]270:$\scalebox{0.7}{}$},
label={[xshift=-0.2cm]90:$\scalebox{0.7}{$F^i_{\alpha_i}$}$}, fill=white](HXP1) at
(7.2,3.6)
{};

\node[inner sep=0](HXPD) at (7.2,2) {$\scalebox{2}{$\vdots$}$};

\draw[dashed] (HXP1) -- ($(HXPD)+(0,0.1)$);
\draw[dashed] (HX) -- (HXPD);

% Y' 

\node[draw, circle, inner sep=2, 
label={[]270:$\scalebox{0.7}{}$},
label={[xshift=-0.2cm]90:$\scalebox{0.7}{$F^{i+1}_{\alpha_{i+1}}$}$}, fill=white](XYP) at (0.9,4.5)
{};

\node[draw, circle, inner sep=2, 
label={[]270:$\scalebox{0.7}{}$},
label={[xshift=-0.2cm]90:$\scalebox{0.7}{$F^{i+1}_{\alpha_{i+1}}$}$}, fill=white](XYP) at
(2.7,4.5)
{};

\node[draw, circle, inner sep=2, 
label={[]270:$\scalebox{0.7}{}$},
label={[xshift=-0.2cm]90:$\scalebox{0.7}{$F^{i+1}_{\alpha_{i+1}}$}$}, fill=white](XYP) at
(5.4,4.5)
{};

\node[draw, circle, inner sep=2, 
label={[]270:$\scalebox{0.7}{}$},
label={[xshift=-0.2cm]90:$\scalebox{0.7}{$F^{i}_{\alpha_i}$}$}, fill=white](XYP) at
(7.2,4.5)
{};

\end{tikzpicture}
\caption{\label{fig:coveredpoints} An intuitive account of the statement of Lemma~\ref{lem:coveredstability}. }
\vspace{-0.2cm}

\end{wrapfigure}
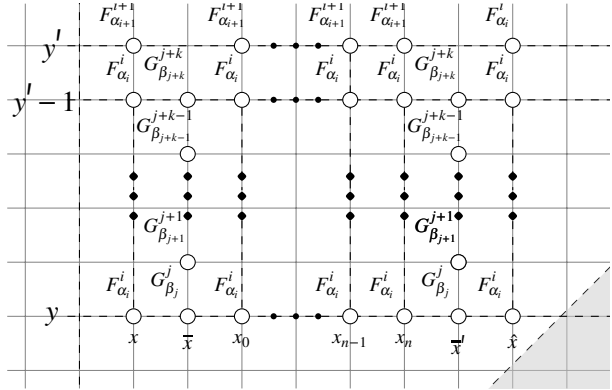

In Figure~\ref{fig:coveredpoints}, we give an intuitive account 
of the notion of covered points and of the statement of 
Lemma~\ref{lem:coveredstability}. First of all, we observe that, 
since $\future(x,y) = \future(x_0,y) = \ldots = \future(x_n,y)$ and, 
for all $0\leq j,j'\leq n$, it holds that $(\shadingGB(x_j), 
\cL(x_j,y))=(\shadingGB(x_{j'}), \cL(x_{j'},y))$, there exists
$x_n <\hx \leq y$ such that $(\shadingGB(x_n), \cL(x_n,y)) = 
(\shadingGB(\hx),$ $\cL(\hx,y))$, and $\hx$ is the smallest 
point greater than $x_n$ that satisfies such a condition. 
Now, it may happen that $\future(x_n,y)  
\supset \future(\hx,y)$, and all points $\ox'> x_n$ with $(\shadingGB(\ox'), \cL(\ox',y)) = (\shadingGB(\ox), \cL(\ox,y))$, 
for some $x < \ox < x_n$, are such that $x_n < \ox' < \hx$.
Then, it can be the case that, for all $0\leq i \leq n$, $\cL(x_i, y')= F^{i+1}_{\alpha_{i+1}}$, 
as all points $(x_i, y')$ satisfy some $D$-request $\psi$  that only belongs to $\cL(\ox', y' - 1)$.
In such a case, as shown in Figure~\ref{fig:coveredpoints}, $\cL(\hx, y') = 
F^{i}_{\alpha_i}$, because for all points
$(\hx', \hy')$, with $\hx <\hx' \leq \hy' < y'$, $\psi \notin \cL(\hx', \hy')$. 
Hence, $(\shadingGB(x_n), F^{i+1}_{\alpha_{i+1}}) \in \future(x_j, y')$ for all $0 \leq j <n$, but $(\shadingGB(x_n), F^{i+1}_{\alpha_{i+1}}) \notin \future(x_n, y')$. Then, by applying Corollary~\ref{cor:inbetweeners}, we have that $\future(x_0, y')= \future(x_{n-1}, y')$. Since
 $\Deltareq(F^{i+1}_{\alpha_{i+1}}) < \Deltareq(F^{i}_{\alpha_{i}}) (=n)$, it holds that $ \Deltareq(F^{i+1}_{\alpha_{i+1}}) \leq  n-1$. 
The same argument can then be applied to  $x, x_0, \ldots, x_{n-1}$ on $y'$, and so on.

%!TEX root = ./main.tex 

\vspace{0.3cm}

\noindent \textbf{Step 4:} A contraction method for homogeneous compass structures.\\

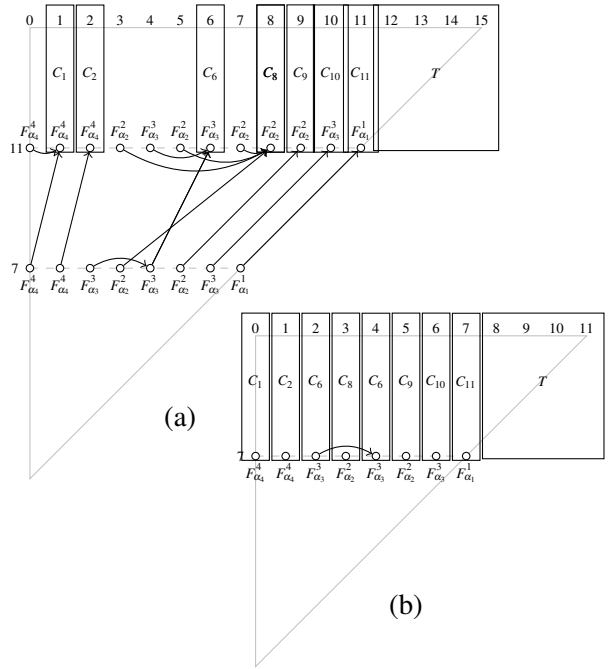
\begin{wrapfigure}{r}{8cm}
    \centering
    \vspace{-0.1cm}
    %!TEX root = ./main.tex 

\begin{tikzpicture}[node distance=0.4cm]

\node[draw, circle, inner sep=1,  fill=white,
label={[label distance=-0.1cm]270:$\scalebox{0.5}{\vpair{
F^4_{\alpha_4} }}$}](0) {};

\node[draw, circle, inner sep=1,  fill=white,
label={[label distance=-0.1cm]270:$\scalebox{0.5}{\vpair{
F^4_{\alpha_4} }}$}, right of=0](1) {};

\node[draw, circle, inner sep=1,  fill=white,
label={[label distance=-0.1cm]270:$\scalebox{0.5}{\vpair{
F^3_{\alpha_3} }}$}, right of=1](2) {};

\node[draw, circle, inner sep=1,  fill=white,
label={[label distance=-0.1cm]270:$\scalebox{0.5}{\vpair{
F^2_{\alpha_2} }}$}, right of=2](3) {};

\node[draw, circle, inner sep=1,  fill=white,
label={[label distance=-0.1cm]270:$\scalebox{0.5}{\vpair{
F^3_{\alpha_3} }}$}, right of=3](4) {};

\node[draw, circle, inner sep=1,  fill=white,
label={[label distance=-0.1cm]270:$\scalebox{0.5}{\vpair{
F^2_{\alpha_2} }}$}, right of=4](5) {};

\node[draw, circle, inner sep=1,  fill=white,
label={[label distance=-0.1cm]270:$\scalebox{0.5}{\vpair{
F^3_{\alpha_3} }}$}, right of=5](6) {};

\node[draw, circle, inner sep=1,  fill=white,
label={[label distance=-0.1cm]270:$\scalebox{0.5}{\vpair{
F^1_{\alpha_1} }}$}, right of=6](7) {};

\node[below of=5, node distance=2cm] {(a)};

\pgftransformshift{\pgfpoint{0cm}{1.6cm}}

\node[draw, circle, inner sep=1,  fill=white,
label={[label distance=-0.1cm]90:$\scalebox{0.5}{\vpair{
F^4_{\alpha_4} }}$}](A0) {};

\node[draw, circle, inner sep=1,  fill=white,
label={[label distance=-0.1cm]90:$\scalebox{0.5}{\vpair{
F^4_{\alpha_4} }}$}, right of=A0](A1) {};

\node[draw, circle, inner sep=1,  fill=white,
label={[label distance=-0.1cm]90:$\scalebox{0.5}{\vpair{
F^4_{\alpha_4} }}$}, right of=A1](A2) {};

\node[draw, circle, inner sep=1,  fill=white,
label={[label distance=-0.1cm]90:$\scalebox{0.5}{\vpair{
F^2_{\alpha_2} }}$}, right of=A2](A3) {};

\node[draw, circle, inner sep=1,  fill=white,
label={[label distance=-0.1cm]90:$\scalebox{0.5}{\vpair{
F^3_{\alpha_3} }}$}, right of=A3](A4) {};

\node[draw, circle, inner sep=1,  fill=white,
label={[label distance=-0.1cm]90:$\scalebox{0.5}{\vpair{
F^2_{\alpha_2} }}$}, right of=A4](A5) {};

\node[draw, circle, inner sep=1,  fill=white,
label={[label distance=-0.1cm]90:$\scalebox{0.5}{\vpair{
F^3_{\alpha_3} }}$}, right of=A5](A6) {};

\node[draw, circle, inner sep=1,  fill=white,
label={[label distance=-0.1cm]90:$\scalebox{0.5}{\vpair{
F^2_{\alpha_2}}}$}, right of=A6](A7) {};

\node[draw, circle, inner sep=1,  fill=white,
label={[label distance=-0.1cm]90:$\scalebox{0.5}{\vpair{
F^2_{\alpha_2} }}$}, right of=A7](A8) {};

\node[draw, circle, inner sep=1,  fill=white,
label={[label distance=-0.1cm]90:$\scalebox{0.5}{\vpair{
F^2_{\alpha_2} }}$}, right of=A8](A9) {};

\node[draw, circle, inner sep=1,  fill=white,
label={[label distance=-0.1cm]90:$\scalebox{0.5}{\vpair{
F^3_{\alpha_3} }}$}, right of=A9](A10) {};

\node[draw, circle, inner sep=1,  fill=white,
label={[label distance=-0.1cm]90:$\scalebox{0.5}{\vpair{
F^1_{\alpha_1} }}$}, right of=A10](A11) {};

\draw[->] (4) -- (A6);
\draw[->] (4) -- (A6);
\draw[->] (5) -- (A9);
\draw[->] (6) -- (A10);
\draw[->] (7) -- (A11);
\draw[->] (1) -- (A2);
\draw[->] (0) -- (A1);
\draw[->] (3) -- (A8);

\draw (A7) edge[bend right, ->] (A8);
\draw (A5) edge[bend right, ->] (A8);
\draw (A4) edge[bend right, ->] (A6);
\draw (A3) edge[bend right, ->] (A8);
\draw (A0) edge[bend right, ->] (A1);

\draw (2) edge[bend left, ->] (4);

\pgftransformshift{\pgfpoint{0cm}{-4.4cm}}

\begin{pgfonlayer}{background}

\draw[opacity=0.25] (0,0) -- (6,6) -- (0,6) -- cycle;
\draw[opacity=0.25, dashed] (0,4.4) -- (4.4,4.4) node[opacity=1, pos = -0.04] {\scalebox{0.7}{$\scalebox{0.7}{11}$}};
\draw[opacity=0.25, dashed] (0,2.8) -- (2.8,2.8) node[opacity=1, pos = -0.07] {\scalebox{0.7}{$\scalebox{0.7}{7}$}};

\node[yshift=0.1cm](L0) at (0,6) {$\scalebox{0.5}{0}$};
\node[right of=L0](L1) {$\scalebox{0.5}{1}$};
\node[right of=L1](L2) {$\scalebox{0.5}{2}$};
\node[right of=L2](L3) {$\scalebox{0.5}{3}$};
\node[right of=L3](L4) {$\scalebox{0.5}{4}$};
\node[right of=L4](L5) {$\scalebox{0.5}{5}$};
\node[right of=L5](L6) {$\scalebox{0.5}{6}$};
\node[right of=L6](L7) {$\scalebox{0.5}{7}$};
\node[right of=L7](L8) {$\scalebox{0.5}{8}$};
\node[right of=L8](L9) {$\scalebox{0.5}{9}$};
\node[right of=L9](L10) {$\scalebox{0.5}{10}$};
\node[right of=L10](L11) {$\scalebox{0.5}{11}$};
\node[right of=L11](L12) {$\scalebox{0.5}{12}$};
\node[right of=L12](L13) {$\scalebox{0.5}{13}$};
\node[right of=L13](L14) {$\scalebox{0.5}{14}$};
\node[right of=L14](L15) {$\scalebox{0.5}{15}$};

\end{pgfonlayer}

\node(T) [below of=L12, node distance =1.6cm] {} ; 

\node[draw, fit=(A1)(L1), inner sep=0] {$\scalebox{0.5}{$C_1$}$};
\node[draw, fit=(A2)(L2), inner sep=0] {$\scalebox{0.5}{$C_2$}$};
\node[draw, fit=(A6)(L6), inner sep=0] {$\scalebox{0.5}{$C_6$}$};
\node[draw, fit=(A8)(L8), inner sep=0] {$\scalebox{0.5}{$C_8$}$};
\node[draw, fit=(A8)(L8), inner sep=0] {$\scalebox{0.5}{$C_8$}$};
\node[draw, fit=(A9)(L9), inner sep=0] {$\scalebox{0.5}{$C_9$}$};
\node[draw, fit=(A10)(L10), inner sep=0] {$\scalebox{0.5}{$C_{10}$}$};
\node[draw, fit=(A11)(L11), inner sep=0] {$\scalebox{0.5}{$C_{11}$}$};
\node[draw, fit=(L12)(L15)(T), inner sep=0] {$\scalebox{0.5}{$T$}$};

\pgftransformshift{\pgfpoint{3cm}{-2.5cm}}

\draw[opacity=0.25] (0,0) -- (4.4,4.4) -- (0,4.4) -- cycle;
\draw[opacity=0.25, dashed] (0,2.8) -- (2.8,2.8) node[opacity=1, pos = -0.07] {\scalebox{0.7}{$\scalebox{0.7}{7}$}};

\node[yshift=0.1cm](L0) at (0,4.4) {$\scalebox{0.5}{0}$};
\node[right of=L0](L1) {$\scalebox{0.5}{1}$};
\node[right of=L1](L2) {$\scalebox{0.5}{2}$};
\node[right of=L2](L3) {$\scalebox{0.5}{3}$};
\node[right of=L3](L4) {$\scalebox{0.5}{4}$};
\node[right of=L4](L5) {$\scalebox{0.5}{5}$};
\node[right of=L5](L6) {$\scalebox{0.5}{6}$};
\node[right of=L6](L7) {$\scalebox{0.5}{7}$};
\node[right of=L7](L8) {$\scalebox{0.5}{8}$};
\node[right of=L8](L9) {$\scalebox{0.5}{9}$};
\node[right of=L9](L10) {$\scalebox{0.5}{10}$};
\node[right of=L10](L11) {$\scalebox{0.5}{11}$};

\pgftransformshift{\pgfpoint{0cm}{2.8cm}}

\node[draw, circle, inner sep=1,  fill=white,
label={[label distance=-0.1cm]270:$\scalebox{0.5}{\vpair{
F^4_{\alpha_4} }}$}](0) {};

\node[draw, circle, inner sep=1,  fill=white,
label={[label distance=-0.1cm]270:$\scalebox{0.5}{\vpair{
F^4_{\alpha_4} }}$}, right of=0](1) {};

\node[draw, circle, inner sep=1,  fill=white,
label={[label distance=-0.1cm]270:$\scalebox{0.5}{\vpair{
F^3_{\alpha_3} }}$}, right of=1](2) {};

\node[draw, circle, inner sep=1,  fill=white,
label={[label distance=-0.1cm]270:$\scalebox{0.5}{\vpair{
F^2_{\alpha_2} }}$}, right of=2](3) {};

\node[draw, circle, inner sep=1,  fill=white,
label={[label distance=-0.1cm]270:$\scalebox{0.5}{\vpair{
F^3_{\alpha_3} }}$}, right of=3](4) {};

\node[draw, circle, inner sep=1,  fill=white,
label={[label distance=-0.1cm]270:$\scalebox{0.5}{\vpair{
F^2_{\alpha_2} }}$}, right of=4](5) {};

\node[draw, circle, inner sep=1,  fill=white,
label={[label distance=-0.1cm]270:$\scalebox{0.5}{\vpair{
F^3_{\alpha_3} }}$}, right of=5](6) {};

\node[draw, circle, inner sep=1,  fill=white,
label={[label distance=-0.1cm]270:$\scalebox{0.5}{\vpair{
F^1_{\alpha_1} }}$}, right of=6](7) {};

\draw (2) edge[bend left, ->] (4);

\node[below of=5, node distance=2cm] {(b)};

\node(T) [below of=L8, node distance =1.6cm] {} ; 

\node[draw, fit=(0)(L0), inner sep=0] {$\scalebox{0.5}{$C_1$}$};
\node[draw, fit=(1)(L1), inner sep=0] {$\scalebox{0.5}{$C_2$}$};
\node[draw, fit=(2)(L2), inner sep=0] {$\scalebox{0.5}{$C_6$}$};
\node[draw, fit=(3)(L3), inner sep=0] {$\scalebox{0.5}{$C_8$}$};
\node[draw, fit=(4)(L4), inner sep=0] {$\scalebox{0.5}{$C_6$}$};
\node[draw, fit=(5)(L5), inner sep=0] {$\scalebox{0.5}{$C_9$}$};
\node[draw, fit=(6)(L6), inner sep=0] {$\scalebox{0.5}{$C_{10}$}$};
\node[draw, fit=(7)(L7), inner sep=0] {$\scalebox{0.5}{$C_{11}$}$};
\node[draw, fit=(L8)(L11)(T), inner sep=0] {$\scalebox{0.5}{$T$}$};

\end{tikzpicture}
    \vspace{-0.4cm}
    \caption{An example of contraction, where compass structure (a) is contracted into compass structure (b).}
    \label{fig:step4picture}
    \vspace{-1.cm}
\end{wrapfigure}

Let us now complete the proof of Theorem \ref{thm:bound2} by providing a small model theorem for compass structures.
By exploiting Lemma \ref{lem:coveredstability}, we can show that, for each row $y$, the cardinality of the set of columns $x_1, \ldots, x_m$ which are not covered
on $y$ is exponential in $|\varphi|$. Then, the sequence of triplets for non-covered points that
appear on $y$ is bounded by an exponential value on $|\varphi|$.

It follows that, in a compass structure of size more than doubly exponential in $|\varphi|$, 
there exist two rows  $y, y'$, with $y < y'$, such that
the sequences of the triplets for non-covered points that
appear on $y$ and $y'$ are exactly the same. 
This allows us to apply a ‘‘contraction'' between $y$  and $y'$ on the compass structure.
An example of how contraction works is given in Figure \ref{fig:step4picture}.

First of all, notice that rows $7$ and $11$ feature the same sequences for triplets of non-covered points, and that, on any row, each covered point is connected by an edge to the non-covered point that ‘‘behaves'' in the same way.
More precisely, we have that column $2$ behaves as column $4$ between $y=7$ and $y'=15$, columns $3, 5$, and $7$ behave as column $8$ between $y=11$ and $y'=15$,
and column $4$ behaves as column $6$ between $y=11$ and $y'=15$.
The compass structure in Figure \ref{fig:step4picture}.(a) can thus be shrinked into the compass structure in Figure \ref{fig:step4picture}.(b), where
each column of non-covered points $x$ on $y'$ is copied 
above the corresponding non-covered point $x'$ on $y$. Moreover, 
the column of a non-covered point $x$ on $y'$ is copied
over all the points which are covered by the non-covered point $x'$ corresponding to $x$ on $y$. This is the case with point $2$ in  Figure \ref{fig:step4picture}.(b)
which takes the new column of its ‘‘covering'' point $4$.
The resulting compass structure is $y'-y$ shorter than the original one, and we can repeatedly apply the contraction step until we achieve the desired bound.

The next corollary, which easily follows from Lemma~\ref{lem:coveredstability},
is crucial for the proof of the EXPSPACE membership of the satisfiability problem for $\mathsf{BDA}_{hom}$. Roughly speaking, it states  that the property of ``being covered'' propagates upward.

\begin{corollary}\label{cor:coveredismonotone}
Let $\cG=(N, \cL)$ be a compass structure. Then, for every covered point $(x,y)$, it holds that, for all $y \leq y' \leq N$, point $(x,y')$ is covered as well. 
\end{corollary}

From Corollary~\ref{cor:coveredismonotone}, it immediately follows that, for every covered point $(x,y)$ and every $y \leq y' \leq N$, there exists $x'> x$ such that $\cL(x',y')=\cL(x,y')$. 
Hence, for all $\ox, \oy$, with $\ox < x \leq y' < \oy$,
and any $D$-request $\psi\in \reqD(\cL(\ox, \oy)) \cap \obsD(\cL(x,y))$, we have that $\psi \in \cL(x',y)$, with $x'>x$. This allows us to conclude that if $(x,y)$ is covered, then all points $(x,y')$, with $y' \geq y$,
are irrelevant from the point of view of $D$-requests.
Let $\cG=(N, \cL)$ be a compass structure and $0\leq y \leq N$.  We define the set of \emph{witnesses} of $y$ as the set 
$\witnesses(y)=\{x: (x,y) \mbox{ is not covered}\}$.
Corollary~\ref{cor:coveredismonotone} guarantees that, for any row $y$, the shading $\shadingGB(x)$ and the labelling $\cL(x,y)$ of witnesses $x \in \witnesses(y)$ are sufficient, bounded, and unambiguous pieces of information that one needs to maintain about $y$. 

Given a compass structure $\cG=(N, \cL)$  and $0\leq y \leq N$, we define the \emph{row blueprint} of $y$ in $\cG$, 
written $\rowG(y)$, as the sequence
 $\rowG(y)= (\shadingB^0, F^0_{\alpha_0})
\ldots (\shadingB^m, F^m_{\alpha_m})$ such that $m + 1 = |\witnesses(y)|$ 
and there exists a bijection $b: \witnesses(y)\rightarrow \{0,\ldots, m\} $ such that, for every $x \in \witnesses(y)$, it holds that $\shadingGB(x) = \shadingB^{b(x)}$ and $\cL(x,y)=F^{b(x)}_{\alpha_{b(x)}}$, and for every $x,x'$ in $\witnesses(y)$, $b(x)<b(x') \leftrightarrow x <x'$. Now, we are ready to prove the following \emph{small model} theorem.

\begin{theorem}\label{thm:smallcompass}
Let $\cG=(N, \cL)$ be a compass structure.  If there exist two points $y, y'$, with $0 \leq y < y'\leq N$, such that
$\rowG(y) = \rowG(y')$, then there exists a compass structure $\cG'=(N', \cL')$ with $N'= N - (y'-y)$.
\end{theorem}

\noindent The proof of Theorem  \ref{thm:bound2}, is completed by proving that if a $\mathsf{BDA}_{hom}$ formula is satisfiable, then it is satisfied by a doubly exponential compass structure, whose existence can be checked in exponential space.

\begin{theorem}\label{thm:bound1}
Let $\varphi$ be a $\mathsf{BD}$ formula. It holds that $\varphi$ is satisfiable iff there is a compass structure $\cG=(N, \cL)$ for it
such that $N \leq 2^{5|\varphi|\cdot(6^{10 |\varphi|^2+4 |\varphi|}\cdot \frac{2}{3}^{10 |\varphi| + 4})}$, whose existence can be checked in $EXPSPACE$.
\end{theorem}

\section{Conclusions}\label{sec:conclusions}

In this paper, we prove that the satisfiability problem for 
$\mathsf{ABD}_{hom}$ over finite homogeneous linear orders 
is EXPSPACE-complete. This result stems a number of observations
regarding the complexity landscape of the satisfiability and 
model checking problems related to $\mathsf{HS}$ interpreted  
over homogeneous structures ($\mathsf{HS}_{hom}$):
\begin{inparaenum}
\item it improves the previously-known non-elementary upper bound 
\cite{DBLP:journals/acta/MolinariMMPP16};
\item it provides a first EXPSPACE-complete fragment of 
$\textsf{HS}_{hom}$ w.r.t. to the satisfiability problem
\cite{DBLP:journals/tcs/BozzelliMMPS19}.
\end{inparaenum}

A more important fact regards how the results for
$\mathsf{ABD}_{hom}$ can enlight
 the problem of determining the exact complexity of the satisfiability problem for the fragment 
 $\mathsf{BE}_{hom}$ which is still 
open today. As a matter of fact $\mathsf{ABD}_{hom}$
and  $\mathsf{BE}_{hom}$ are not comparable from an expressive 
standpoint \cite{DBLP:journals/tcs/BresolinMMSS14}.
However, by means of $\mathsf{ABD}_{hom}$ we can capture 
a fragment of $\mathsf{BE}_{hom}$ that is $\mathsf{BD}_{hom}$
plus a restricted version of the $\hsEE$ operator
namely $\hsEE_{\pi} \psi = \hsEA(\pi \wedge \psi)$
that allows one to predicate on the ending point of an interval.
As we show in Section~\ref{sec:hardness}, this 
is the only key property that enables the jump in complexity
from $\mathsf{BD}_{hom}$ (PSPACE-Complete)
to $\mathsf{ABD}_{hom}$ (EXPSPACE-Complete)
w.r.t. the satisfiability problem. 
It is easy to see that the result presented here 
can be easily extended to the case of homogeneous 
structures isomorphic to $\bbN$.

In the future we plan to consider the satisfiability/model checking problem of (fragments of) $\mathsf{HS}_{hom}$ interpreted over linear order like $\bbQ$
and $\bbR$. Finally, let us point out that 
the precise characterization of the complexity of the satisfiability problem for $\mathsf{BE}_{hom}$ over finite structures 
is still the main open problem on the path of 
determining the complexity of the satisfiability problem for $\mathsf{HS}_{hom}$.

\bibliographystyle{eptcs}
\bibliography{biblio}
\end{document}